\documentclass[lettersize,journal]{IEEEtran}
\usepackage{cite}
\usepackage{amsmath,amssymb,amsfonts}
\usepackage{algorithmicx}
\usepackage[noend,ruled, linesnumbered, vlined]{algorithm2e}

\usepackage{array}
\usepackage{graphicx}
\usepackage{stfloats}
\usepackage{amsthm}
\usepackage{subfigure}
\usepackage{url}
\usepackage{verbatim}
\usepackage{algpseudocode}  
\usepackage{textcomp}
\usepackage{xcolor}
\usepackage{color}

\newenvironment{sequation}{\begin{equation}\small}{\end{equation}}

\usepackage{multirow}
\usepackage{bm}
\usepackage{booktabs}

\usepackage{epstopdf}
\usepackage{cases}
\usepackage{makecell,multirow,diagbox}
\usepackage{stfloats}
\usepackage{comment}
\usepackage{array}
\usepackage{xr}

\definecolor{b}{rgb}{0, 0, 0}
\definecolor{k}{rgb}{0, 0, 0}
\definecolor{b1}{rgb}{0, 0, 0}

\begin{document}

\title{Task Offloading and Resource Allocation for MEC-assisted Consumer Internet of Vehicle Systems}

\author{
Yanheng~Liu, 
Dalin Li, 
Hao Wu,
Zemin~Sun,~\IEEEmembership{Member,~IEEE}, 
Weihong Qin, 
Jun~Li, 
Hongyang~Du, and 
Geng~Sun,~\IEEEmembership{Senior Member,~IEEE}
\thanks{This study is supported in part by the National Natural Science Foundation of China (62172186, 62272194, 62471200), in part by the Science and Technology Development Plan Project of Jilin Province (20250101027JJ), in part by the Engineering Technology Research Center of Guangdong Provincial Key Scientific Research Platform Project (2024GCZX001), in part by Guangdong Provincial Key Disciplines Project (2024ZDJS142, 2022ZDJS139), in part by Zhuhai Industry University Research Project (2220004002671), and in part by China Postdoctoral Fund (2023M731282). (\textit{Corresponding author: Zemin Sun.}).
\par Yanheng Liu, Dalin Li, and Hao Wu are with the School of Computer Science, Zhuhai College of Science and Technology, Zhuhai 519041, China. (e-mail: yhliu@jlu.edu.cn; lidalin@zcst.edu.cn; wuhao@zcst.edu.cn).
\par Zemin Sun is with the College of Computer Science and Technology, Key Laboratory of Symbolic Computation and Knowledge Engineering of Ministry of Education, Jilin University, Changchun 130012, China (e-mail: sunzemin@jlu.edu.cn).
\par Weihong Qin is with the College of Software Engineering, Jilin University, Changchun 130012, China. (e-mail: qinwh23@mails.jlu.edu.cn).
\par Jun Li is with the School of Electronics and Communication Engineering, Guangzhou University, Guangzhou 510006, China, (email: lijun52018@gzhu.edu.cn).
\par Hongyang Du is with the Department of Electrical and Electronic Engineering, University of Hong Kong, Pok Fu Lam, Hong Kong (e-mail: duhy@eee.hku.hk).
\par Geng Sun is with the College of Computer Science and Technology, Key Laboratory of Symbolic Computation and Knowledge Engineering of Ministry of Education, Jilin University, Changchun 130012, China, and also with the College of Computing and Data Science, Nanyang Technological University, Singapore 639798 (e-mail: sungeng@jlu.edu.cn).
}
}

\IEEEtitleabstractindextext{
\begin{abstract}
\par Mobile edge computing (MEC)-assisted internet of vehicle (IoV) is emerging as a promising paradigm to provide computing services for vehicles. However, meeting the computing-sensitive and computation-intensive demands of vehicles poses several challenges, including the discrepancy between the limited resource provision and stringent computing requirement, the difficulty in capturing and integrating the intricate features of the MEC-assisted IoV system into the problem formulation, and the need for real-time processing and efficient resource management in the dynamic environment. In this work, we explore the AI-enabled task offloading and resource allocation for MEC-assisted consumer IoV systems. Specifically, we first present a multi-MEC-assisted consumer IoV architecture that leverages the computational resources of MEC servers to provide offloading services close to vehicles. Subsequently, we formulate a system cost minimization optimization problem (SCMOP) by integrating the service delay and energy consumption. To efficiently solve this problem, we design a joint task offloading and computing resource allocation approach (JTOCRA) by applying the multi-agent deep deterministic policy gradient (MADDPG) algorithm. Finally, simulation results demonstrate that the proposed JTOCRA can achieve superior system performances and exhibits better scalability compared to other alternative approaches.

%The convergence of artificial intelligence (AI), multi-access edge computing (MEC), and consumer internet of vehicle (IoV) is emerging as a critical technology for supporting the development of future intelligent transportation systems. In this context, we present a study on AI-enabled task offloading and resource allocation for MEC-assisted consumer IoV systems to drive the advancement of various intelligent vehicular applications. 
\end{abstract}
	
}

\maketitle
\IEEEdisplaynontitleabstractindextext
\IEEEpeerreviewmaketitle

%	
% Introduction
%

\section{Introduction}
\label{sec_introduction}

\par \IEEEPARstart{B}{enefiting} from advancements in artificial intelligence (AI) and the Internet of Things (IoT), the advent of next-generation consumer electronics has ushered in a new era of technological innovation, characterized by enhanced connectivity, advanced sensor capabilities, and increased computational power. These advancements not only revolutionize personal devices but also pave the way for more sophisticated and integrated systems across various domains. One significant area benefiting from these innovations is the consumer internet of vehicles (IoV), where vehicles extend the functionality of traditional consumer electronics by embedding them within the vehicular networks. In the consumer IoV, vehicles function as consumer electronics and are connected through IoV frameworks to create a more connected, efficient, and intelligent transportation system. 

\par The proliferation of vehicles on the road, coupled with advancements in the sixth generation (6G) communication, are propelling the development of various intelligent vehicular applications such as \textcolor{b}{in-vehicle entertainment, autonomous navigation, and collision detection~\cite{WangWGQH24,AbbasFK19}}. Most of these applications, particularly those related to autonomous driving and safety, are mission-critical, leading to \textcolor{b}{stringent requirements for computing resources and low latency~\cite{HouHZLHC24}}. Moreover, the rapid development of generative AI further facilitates new applications such as human-machine interaction~\cite{Sun2025GenAI, He2025GAI, Xie2024GenerativeAF}. However, the limited computing capabilities of consumer vehicles present a significant challenge in handling these {\textcolor{b}delay-sensitive and computation-intensive tasks~\cite{ErnestM24}}. To address this challenge, mobile edge computing (MEC) offers a promising solution by migrating the cloud capabilities to road side units (RSUs) located in \textcolor{b}{close proximity to vehicles~\cite{Liu2025}}. Consequently, by offloading the task to MEC servers, the workload on resource-constrained vehicles can be alleviated, and the latency-critical tasks can be processed timely.

\par Despite the abovementioned benefits, designing an efficient approach for task offloading in the MEC-assisted consumer IoV system presents several challenges. \textit{\textbf{First}}, compared to cloud servers, MEC server is equipped with limited computing and energy resources. However, the proliferated delay-sensitive and computation-hungry tasks pose stringent requirements on edge computing services. Therefore, a single MEC server may struggle to handle the massive and stringent task demands from consumer vehicles, particularly in dense areas or during peal hours. When a large number of tasks are offloaded simultaneously, the single MEC server could be overloaded or congested, leading to a degradation in service quality. \textit{\textbf{Second}},  different from conventional consumer electronics systems, consumer IoV systems exhibit intricate attributes such as the high mobility of vehicles, the dynamic and unstable nature of wireless networks, the stochastic generation of tasks, and the random distribution of vehicles. Additionally, different nodes within the system, including vehicles and MEC servers, have heterogeneous preferences regarding system performance metrics such as latency and energy consumption. Accurately capturing and integrating these intricate and diverse features into the problem formulation poses a significant challenge. \textit{\textbf{Third}}, the mission-critical nature of most vehicular tasks imposes strict requirements on real-time task processing, while the limited resources of MEC servers necessitate cost-effective and energy-efficient resource management. Moreover, the highly dynamic nature of the MEC-assisted consumer IoV system such as the mobility of vehicles, time-varying wireless channel, and random arrival of tasks, add complexity to designing an efficient approach of task offloading and computing resource allocation. Such approach should be capable of meeting the delay-sensitive demands for vehicles, ensuring the computing resource constraints and energy efficiency for MEC servers, as well as adapting to the dynamic of the system.

\par To address the abovementioned challenges, we present a joint optimization approach of task offloading and computing resource allocation for the MEC-assisted consumer IoV system. The main contributions are summarized as follows:

\begin{itemize}
      \item \textit{\textbf{Edge Computing Architecture.}} We consider a multi-MEC-assisted consumer IoV architecture, where multiple MEC servers are deployed on the RSUs close
      to vehicles for task processing. Specifically, multiple MEC servers can effectively alleviate the constraints of a single MEC in terms of computing resources, energy capacity, and coverage range, thereby expanding the responsiveness and applicability of the system.
      
      \item \textit{\textbf{Problem Formulation for System Cost Minimization.}} We formulate a system cost minimization optimization problem (SCMOP) by jointly optimizing the task offloading and computing resource allocation. Specifically, the system cost theoretically modeled by synthesizing the vehicle mobility, channel characteristics, service delay, and energy consumption. Moreover, the SCMOP is proved as a non-convex and NP-hard mixed integer nonlinear programming (MINLP) problem.
      
      \item \textit{\textbf{Joint Optimization Algorithm Design.}} To solve SCMOP, we propose a joint task offloading and computing resource allocation approach (JTOCRA). Specifically, we first reformulate the SCMOP into a Markov decision game to enhance its portability. Moreover, we present a multi-agent deep deterministic policy gradient (MADDPG)-based algorithm to determine the decisions of task offloading and computing resource allocation.
	
      \item \textit{\textbf{Effectiveness and Efficiency Validation.}} We evaluate the performance of JTOCRA through simulation. Specifically, simulation results demonstrate that the proposed JTOCRA is able to achieve superior performance in terms of the system cost and task completion delay while meeting the energy constraint. We also find that the proposed JTOCRA exhibits better scalability in the considered scenario.
\end{itemize}

\par The rest of this paper is organized as follows. Section \ref{sec_related work} reviews the related work. Section \ref{sec_system_model} presents the system model. Section \ref{sec_problem_analysis} gives the problem formulation and analysis. Section \ref{sec_jointOffloading} elaborates the proposed approach. Section \ref{sec_simulation} showcases the simulation results. Finally, the conclusions are drawn in Section \ref{sec_conclusion}.

%
% Related work
%
\section{Related work}
\label{sec_related work}
\par In this section, we provide a comprehensive review of existing research focusing on MEC-assisted consumer IoV computing architecture, optimization problem formulation, and optimization approaches. Furthermore, we analyze the limitations of existing research and highlight the novelty of this work.

\subsection{MEC-assisted Consumer IoV Computing Architecture}
\par A lot of research has been conducted to explore effective computing architecture for MEC-assisted consumer IoV systems. Specifically, some works studied computing architectures based on a single MEC server. For example, Wang \emph{et al.}~\cite{wang2024amtos} proposed a four-tier computing architecture, where an edge MEC server, a remote mobile cloud computing server and multiple idle vehicles collaborate to provide computing services for task vehicles. Liu \emph{et al.}~\cite{LiuYZCYT23} designed a single MEC server-assisted IoV computing architecture and proposed a task offloading and resource allocation scheme to reduce the average latency of tasks. However, a single MEC server has limited computational resources and energy supply, which makes it difficult to meet the computational demands of consumer vehicles in large-scale IoV scenarios, particularly during peak periods.

\par The computing architecture based on multiple MEC servers, which can effectively mitigate the limitations of a single MEC server, is gaining widespread attention. For example, Wang \emph{et al.}~\cite{Wang2024} investigated the integration of multiple MEC servers to handle dependent tasks in the IoV. Ning \emph{et al.}~\cite{NingZ00000K21} conducted a study on intelligent computation offloading and content caching in a multi-MEC server-assisted IoV system. Jia \emph{et al.}~\cite{JiaZHZ22} studied computation offloading in a multi-vehicle IoV system assisted by multiple MEC servers. However, the above works usually assume that the computing requirements of vehicles are known or fixed in the considered architecture, which may not accurately reflect the physical characteristics of real IoV systems. Different from the above works, in this work, we propose a multi-MEC server-assisted consumer IoV computing architecture where vehicles have time-varying and heterogeneous computing requirements.

\subsection{Optimization Problem Formulation}

\par Due to the limited computational resources and energy supply of MEC servers, precise formulation of optimization problems is crucial for enhancing the performance of MEC-assisted consumer IoV systems. For example, Zhu \emph{et al.}~\cite{ZhuLLBZ21} formulated a computation offloading problem to minimize the processing delay of all tasks. Huang \emph{et al.}~\cite{Huang2022} jointly optimized task offloading and resource allocation to reduce the energy consumption of vehicles. Zhang \emph{et al.}~\cite{ZhangGLZ20} proposed an software-defined networking based task offloading architecture in which task offloading is optimized to minimize the task processing latency. To minimize system energy consumption, Wang \emph{et al.}~\cite{WangNGW22} conducted research on task offloading for consumer vehicles and resource allocation for MEC server. 

\par Although the above works have enhanced the performance of MEC-assisted consumer vehicular networks, there are some limitations. Specifically, due to the time-varying and computational demands of vehicles, coupled with the constrained computational resources of MEC servers, task offloading and resource allocation are critical factors in improving system performance. Moreover, given the computation-intensive and delay-sensitive nature of vehicular applications, latency and energy consumption are crucial metrics for evaluating system performance. The above studies have not comprehensively considered these factors. Therefore, we formulate a joint task offloading and resource allocation optimization problem to minimize the system cost to make up for the shortcomings of the above studies, where the system cost is constructed by jointly considering latency and energy consumption.

\subsection{Optimization Approach}
\par To tackle the intricate joint optimization problems of task offloading and resource allocation, numerous studies have developed effective algorithms based on traditional methods, such as heuristic algorithms~\cite{Laboni2024Hyper}, swarm intelligent algorithms~\cite{Souza2023}, and convex algorithm~\cite{Men2024}. For example, Yadav~\cite{Yadav2020} presented a heuristic algorithm to jointly optimize the energy and latency for MEC-assisted vehicular network. Moreover, Souza \emph{et al.}~\cite{Souza2023} proposed a bee colony-based algorithm to optimize the task offloading time in vehicular edge computing systems. Men \emph{et al.}~\cite{Men2024} employed the convex optimization algorithm for resource allocation of the vehicular network. However, these traditional algorithms have high computational complexity due to relying on theoretical models, which are not suitable for the dynamic vehicular scenario.

\par Artificial intelligence (AI) algorithms such as deep reinforcement learning (DRL) are gaining widespread attention as a remarkable strategy. For example, Xu \emph{et al.}~\cite{li2023flexedge} introduced a proximal policy optimization (PPO)-based algorithm for task offloading and uncrewed aerial vehicle (UAV) trajectory planning in the UAV-assisted vehicular network. Kong \emph{et al.}~\cite{KongDHSWYC22} proposed a joint caching and computation framework by integrating deep deterministic policy gradient (DDPG) algorithm to reduce the energy expenditure of network service providers. \textcolor{b}{Moreover, Hazarika \emph{et al.}~\cite{Hazarika2024} designed a multi-agent DRL framework for optimizing the task offloading for vehicles in multiple reconfigurable intelligent surfaces (RIS)-aided IoV networks. \textcolor{b1}{Furthermore, Xiao \emph{et al.}~\cite{Xiao2024Joint} proposed a DRL-based approach for UAV trajectory planning and data collection in UAV-assisted IoT networks.} Additionally, Zhao \textit{et al.}~\cite{Zhao2025} studied a multi-hop task offloading scheme using the asynchronous advantage actor-critic (A3C) method, which addresses the issue of unreliable connectivity between RSUs and vehicles. Besides, Chen \textit{et al.}~\cite{Chen2025Veh} presented a two-stage DRL approach to address the communication resource allocation and load balance in vehicular edge computing (VEC) networks.} However, these works did not consider the interactions among different nodes for decision making, which could result in sub-optimal outcomes. Consequently, they are not suitable for the multi-vehicle and dynamic IoV system where the decisions of vehicles are mutually coupled.

% 返修增加：
% [1] Multi-Objective Optimization for UAV Swarm-Assisted IoT with Virtual Antenna Arrays, IEEE Transactions on Mobile Computing, 23(5): 4890-4907, 2024
% [2] Joint Task Offloading and Resource Allocation in Aerial-Terrestrial UAV Networks with Edge and Fog Computing for Post-Disaster Rescue, IEEE Transactions on Mobile Computing, 23(9): 8582-8600, 2024
% [3] BARGAIN-MATCH: A Game Theoretical Approach for Resource Allocation and Task Offloading in Vehicular Edge Computing Networks, IEEE Transactions on Mobile Computing, 23(2): 1655-1673, 2024.
% [4] Collaborative Ground-Space Communications via Evolutionary Multi-objective Deep Reinforcement Learning, IEEE Journal on Selected Areas in Communications, 2024.
% [5] Multi-objective Optimization for Multi-UAV-assisted Mobile Edge Computing, IEEE Transactions on Mobile Computing, 2024
% [6] J Wang, H Du, Y Liu, G Sun, D Niyato, S Mao, DI Kim, X Shen, Generative AI based Secure Wireless Sensing for ISAC Networks. arXiv preprint arXiv:2408.11398.
% [7] Wang, Jiacheng, \emph{et al.} "Generative AI Enabled Robust Data Augmentation for Wireless Sensing in ISAC Networks." arXiv preprint arXiv:2502.12622 (2025).
% [8] Split Federated Learning for UAV-Enabled Integrated Sensing, Computation, and Communication[J]. arXiv preprint arXiv:2504.01443, 2025.

%
% System Model
%
\section{System Model}
\label{sec_system_model}

\par In this section, we first introduce the proposed MEC-assisted consumer IoV system. Then, we elaborate on the relevant basic models, communication models, computational models, and performance metrics.

\begin{table*}[!hbp]
        \setlength{\abovecaptionskip}{-1em}%    
	\setlength{\belowcaptionskip}{0pt}%
	\caption{\textcolor{b}{Summary of notations}}
	\label{tab_notation}
	\renewcommand*{\arraystretch}{0.9}
	\begin{center}
		\begin{tabular}{|p{.12\textwidth}|p{.33\textwidth}||p{.12\textwidth}|p{.33\textwidth}|}
        \hline
            Symbol&Description&Symbol&Description\\
        \hline
            $\mathcal{V}$&The set 
            of consumer vehicles&$T$&Time slots\\
        \hline
             $\mathcal{M}$& The set of MEC servers&$\mathbf{\Psi}$& The task\\
        \hline
             $F_v$&The computing resources&$E_v$&The energy constraint\\
        \hline
             $\mathbf{q}_v^t$&The position of vehicle $v$&$l_v^t$&the data size of the task (in bit)\\
        \hline
             $\mu_v^t$&The computation intensity of the task&$\tau_v^t$&The deadline of the task\\
        \hline
             $\mathbf{St}_m$&The attributes of each MEC server&$F_m$&The computing resources of MEC server $m$\\
        \hline
             $E_m$&The energy constraint of MEC server $m$&$\mathbf{q}_m$&The location of MEC server $m$\\
        \hline
             $\mathbf{v}_v^t$&The velocity of vehicle $v$ at time slot $t$&$\omega\in[0,1]$&The memory degree\\
        \hline
             $\sigma$&The asymptotic standard deviation of velocity&$\bar{\mathbf{v}}_v$&The asymptotic mean of velocity\\
        \hline
             $\mathbf{q}_{v}^{t+1}$&The position of vehicle $v$ at time slot $t+1$&$B_{v,m}$&the subchannel bandwidth \\
        \hline
             $p_{v}^{t}$&The transmit power of vehicle $v$ at time slot $t$&$N_0$&The background noise power\\
        \hline
             $g_{v,m}^t$&The instantaneous channel power gain&$\mathbb{P}_{v,m}^t$&The LoS probability of the communication channel \\
        \hline
             $g_{v,m}^{\mathbf{x},t}$&The channel power gain for LoS or NLoS communication&$d_{v,m}^t$&The distance between vehicle $v$ and MEC server $m$ \\
        \hline
             $\alpha_1$ and $\alpha_2$&The environment-dependent parameters&$g_{v,m}^{\mathbf{x},t}$&The channel power gain \\
        \hline
             $h_{v,m}^{\mathbf{x},t}$&The parameters of small-scale fading for LoS or NLoS communications&$L_{v,m}^{\mathbf{x},t}$&The parameters of large-scale fading for LoS or NLoS communications \\
        \hline
             $\textbf{m}^\mathbf{x}$&The shaping factor of the Nakagami fading channel for LoS or NLoS communication&$\overline{p}$&The average received power \\
        \hline
             $\Gamma(\cdot)$&The Gamma function&$\bar{d}$&The reference path loss for unit distance \\
        \hline 
             $f_c$&The carrier frequency&$c$&The light speed \\
        \hline
             $\beta^\mathbf{x}$&The path loss exponent for LoS or NLoS communication&$\chi^\mathbf{x}$&The shadowing attenuation for LoS or NLoS communication \\
        \hline
             $k_{v,z}^t\in\{0,1\}$&The task offloading decision&$\mathcal{M}\cup v$&The set of feasible destinations for vehicle $v$ to offload task $\mathbf{\Psi}_v^t$ \\
        \hline
             $D_{v,v}^t/D_{v,m}^t$ &The service delay for local/edge&$E_{v,v}^t/E_{m,v}^t$&The energy consumption of vehicle for local/edge \\
        \hline
             $D_v^t$&The service delay for completing task $\mathbf{\Psi}_v^t$&$E_v^t$&The energy consumption for completing task $\mathbf{\Psi}_v^t$ \\
        \hline
             $C_v^t$&The system cost of completing task $\mathbf{\Psi}_v^t$ at time slot $t$&$\mathbf{K}$&The strategies of task offloading \\
        \hline
             $\mathbf{F}$&The computing resource allocation&$\mathcal{S}$&The set of all possible states of the environment \\
        \hline
             $\mathcal{N}$&The set of agents&$\mathcal{O}$&The set of observations available to each agent \\
        \hline
             $\mathcal{A}$&The actions available to the agents&$\mathcal{P}$&The state transition probability function \\
        \hline
		\end{tabular}
	\end{center}
\end{table*}

\subsection{System Overview}
\label{subsec:system_overview}

\par
\begin{figure}[!hbt] 
	\centering
        \setlength{\abovecaptionskip}{2pt}%    
	\setlength{\belowcaptionskip}{2pt}%
	\includegraphics[width =3.4in]{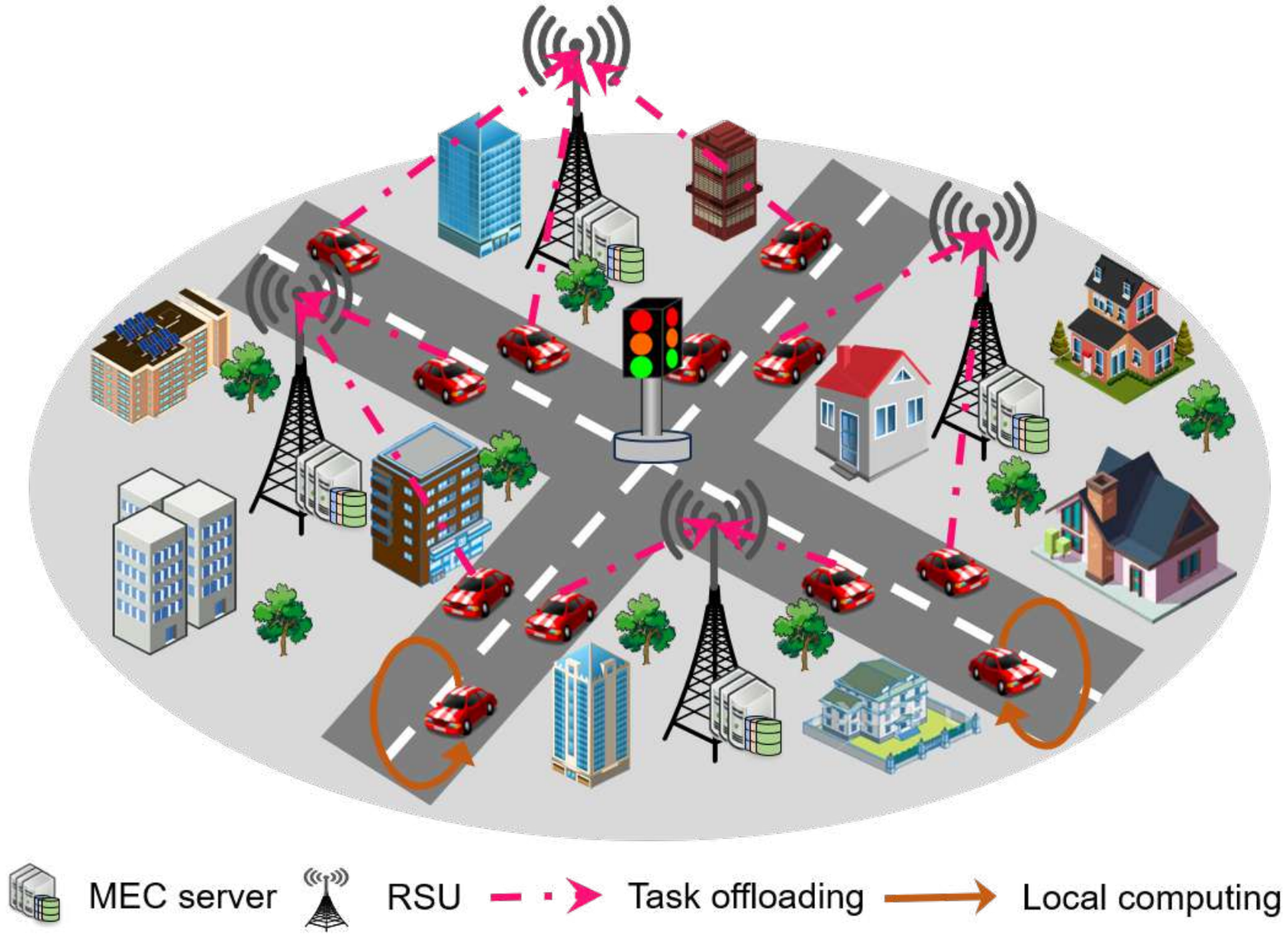}
	\caption{The architecture of MEC-assisted consumer IoV communications.}
	\label{fig_systemModel}
    \vspace{-1.5em}
\end{figure}

 As shown in Fig.~\ref{fig_systemModel}, we consider \textcolor{b}{an MEC}-assisted consumer IoV system, which consists of a set of consumer vehicles $\mathcal{V}=\{1,\ldots,v,\ldots, V\}$ and a set of MEC servers mounted on BSs $\mathcal{M}=\{1,\ldots,m,\ldots, M\}$ \footnote{Note that the BS and the MEC server will be used interchangeably}. Additionally, the consumer IoV system operates in a time-slotted manner, with the system timeline discretized into $T$ time slots $\mathcal{T}=\{1,\ldots,t,\ldots,T\}$, each of equal duration $\delta$.

% Basic models
\subsection{Basic Models}
\label{sec:basic models}

\par The basic models describe the state information of  various entities in the proposed system.

\par \textit{\textbf{Vehicle Model.}} We consider that each vehicle generates one computing task per time slot~\cite{Sun2024Joint}. The attributes of each vehicle $v\in\mathcal{V}$ at time slot $t$ can be characterized as $\mathbf{St}_v(t)=\left(\mathbf{\Psi}_v^t, F_v, E_v, \mathbf{q}_v^t\right)$, where $\mathbf{\Psi}_v^t$ denotes the task generated by vehicle $v$ at time slot $t$, $F_v$ denotes the computing resources of vehicle $v$ (in cycles/s), $E_v$ represents the energy constraint of vehicle $v$, and $\mathbf{q}_v^t=[x_v^t,y_v^t]$ means the position of vehicle $v$ at time slot $t$. In specific, the task $\mathbf{\Psi}_v^t$ is characterized as $\mathbf{\Psi}_v^t=(l_v^t, \mu_v^t, \tau_v^t)$, where $l_v^t$ represents the data size of the task (in bit), $\mu_v^t$ indicates the computation intensity of the task (in cycles/bit), and $\tau_v^t$ denotes the deadline of the task. 

\par \textit{\textbf{MEC Server Model.}} We consider that each MEC server is equipped with multiple CPU cores for parallel task processing. The attributes of each MEC server $m\in\mathcal{M}$ can be characterized as $\mathbf{St}_m=\left(F_m, E_m, \mathbf{q}_m\right)$, where $F_m$ denotes the computing resources of MEC server $m$, $E_m$ represents the energy constraint of MEC server $m$, and $\mathbf{q}_m=[x_m,y_m]$ means the location of MEC server $m$.

\par \textit{\textbf{Vehicle Mobility Model.}} To capture the temporal-dependent movement of vehicles, we employ the Gauss Markov model~\cite{liang1999predictive} to model the mobility of vehicles. In specific, the velocity of vehicle is updated as:
\begin{equation}
\begin{aligned}
	\label{eq_MD_mobility}
		& \mathbf{v}_{v}^{t+1}=\omega\mathbf{v}_{v}^t+(1-\omega) \bar{\mathbf{v}}_v+\sqrt{1-\omega^2} \mathbf{w}_{v}^t, \, \forall v\in\mathcal{V}, \, t\in\mathcal{T},
\end{aligned}
\end{equation}

\noindent where $\mathbf{v}_v^t$ denotes the velocity of vehicle $v$ at time slot $t$. Moreover, $\omega\in[0,1]$ represents the memory degree, which indicates the temporal-dependent level of the velocity. In addition, $\mathbf{w}_v^t$ indicates the uncorrelated random Gaussian process, i.e., $ \mathbf{w}_v^t\sim f^{\text{Gua}}\left(0,\sigma^2\right)$, where $\sigma$ means the asymptotic standard deviation of velocity. Besides,  $\bar{\mathbf{v}}_v$ is the asymptotic mean of velocity. Therefore, the instantaneous position of each vehicle can be updated as:
\begin{equation}
	\label{eq_MD_position}
	\begin{aligned}	\mathbf{q}_{v}^{t+1}=\mathbf{q}_{v}^t + \mathbf{v}_v^t\delta, \, \forall v\in \mathcal{V}, \,  t \in \mathcal{T},
	\end{aligned}
\end{equation}

\noindent where $\mathbf{q}_{v}^{t+1}$ represents the position of vehicle $v$ at time slot $t+1$.

%
%Communication model
%
\subsection{Communication Models}
\label{sec_communicationModel}

\par We consider that the frequency band is orthogonally allocated to each MEC server to mitigate unreliable transmissions caused by interference among different MEC servers. Furthermore, we employ the widely used orthogonal frequency division multiple access (OFDMA) scheme to enable each MEC sever to offer computing service for multiple vehicles simultaneously. Accordingly, the instantaneous transmission data rate between vehicle $v\in\mathcal{V}$ and MEC server $m\in\mathcal{M}$ can be calculated as
\begin{equation}
	\label{eq_dataRate}
	R_{v,m}^t=B_{v,m} \log_2 \big(1+p_{v}^{t} g_{v,m}^t/N_0\big),
\end{equation}

\noindent where $B_{v,m}$ denotes the subchannel bandwidth between vehicle $v$ and MEC server $m$, $p_{v}^{t}$ represents the transmit power of vehicle $v$ at time slot $t$, $N_0$ is the background noise power, and $g_{v,m}^t$ indicates the instantaneous channel power gain between vehicle $v$ and MEC server $m$, which is detailed as follows.

\par The wireless links of consumer IoV communication could experience scattering or shadowing due to the complex and uncertain IoV environment such as occasional obstacles. Therefore, we model the channel power gain by adopting the widely used probabilistic line-of-sight (LoS) channel along with the large-scale and small-scale fadings as follows:
\begin{equation}
	\label{eq_dataRate}
       g_{v,m}^t={\mathbb{P}_{v,m}^t} g_{v,m}^{\text{L}, t} +(1-\mathbb{P}_{v,m}^t)g_{v,m}^{\text{N}, t}, 
\end{equation}

\noindent where $\mathbb{P}_{v,m}^t$ represents the LoS probability of the communication channel, $\mathbf{x}\in\{\text{L},\text{N}\}$) collectively denotes the LoS or non-line-of-sight (NLoS) communication link, and $g_{v,m}^{\mathbf{x},t}$ means the channel power gain for LoS or NLoS communication. Specifically, $\mathbb{P}_{v,m}^t$ and $g_{v,m}^{\mathbf{x},t}$ are detailed as follows.

\subsubsection{LoS Probability}

\par According to the 3GPP standard~\cite{{3GPPTR389012020}}, the LoS probability for vehicle-to-infrastructure (V2I) link can be calculated as
\begin{equation}
\label{eq_LoS_probability_MBS}
    \mathbb{P}_{v,m}^t=\min \big(\alpha_1/d_{v,m}^t, 1\big)\big(1-e^{-\frac{d_{v,m}^t}{\alpha_2}}\big)+e^{-\frac{d_{v,m}^t}{\alpha_2}},
\end{equation}

\noindent where $d_{v,m}^t=\|\mathbf{q}_v^t-\mathbf{q}_m\|$ represents the distance between vehicle $v$ and MEC server $m$ at time slot $t$. Moreover, $\alpha_1$ and $\alpha_2$ denote the environment-dependent parameters. 

\subsubsection{Channel Power Gain}

\par The channel power gain between vehicle $v$ and MEC server $m$ can be calculated as~\cite{Sun2025TJCCT}
\begin{equation}
\label{eq_channel_power_gain}   g_{v,m}^{\mathbf{x},t}=|h_{v,m}^{\mathbf{x},t}|^2\left(L_{v,m}^{\mathbf{x},t}\right)^{-1}, \, \mathbf{x}\in\{\text{L},\text{N}\},
\end{equation}

\noindent where $h_{v,m}^{\mathbf{x},t}$ and $L_{v,m}^{\mathbf{x},t}$ represent the parameters of small-scale fading and large-scale fading for LoS or NLoS communications, respectively. \textcolor{b}{First, we employed  Nakagami-$m$ fading channel to model the small-scale fading since it is well-suited and parametric-scalable for describing different fading scenarios by adjusting the shaping factor of the Nakagami fading channel ${m}$~\cite{ji2020research}. For example, Rayleigh fading is a special case of Nakagami-$m$ fading channel when ${m}=1$. Therefore, the small-scale fading is modeled by applying the Nakagami-$m$ fading as}
\begin{equation}
\label{eq_nakagami}
\begin{aligned}
    &h_{v,m}^{\mathbf{x},t}\sim f^{\text{Nak}}\big(h_{v,m}^{\mathbf{x},t},\textbf{m}^\mathbf{x}\big)\\
    & = \frac{2{(\mathbf{m}^\mathbf{x})}^{\mathbf{m}^\mathbf{x}} ({h_{v,m}^{\mathbf{x},t}})^{2\mathbf{m}^\mathbf{x}-1} e^{\big(-\frac{\mathbf{m}^\mathbf{x}  ({h_{v,m}^{\mathbf{x},t}})^2}{\overline{p}}\big)}}{\Gamma(\mathbf{m}^\mathbf{x}) (\overline{p})^{\textbf{m}^\mathbf{x}}},
\end{aligned}
\end{equation}

\noindent where $\textbf{m}^\mathbf{x}\geq0.5$ denotes the shaping factor of the Nakagami fading channel for LoS or NLoS communication, $\overline{p}$ represents the average received power, and $\Gamma(\cdot)$ is the Gamma function.

\par Moreover, the large-scale fading is calculated by combining the pathloss fading and shadowing, i.e., 
\begin{equation}
\label{eq_nakagami}   
    L_{v,m}^{\mathbf{x},t}=(4\pi  \bar{d} f_c)^2/c^2 \big(d_{v,m}^t/\bar{d}\big)^{\beta^\mathbf{x}} \chi^\mathbf{x},\, 
\end{equation}

\noindent where $\bar{d}$ represents the reference path loss for unit distance, $f_c$ denotes the carrier frequency, $c$ is the light speed, $\beta^\mathbf{x}$ means the path loss exponent for LoS or NLoS communication, and $\chi^\mathbf{x}$ indicates the shadowing attenuation for LoS or NLoS communication, which typically modeled as a random variable following zero-mean Gaussian distribution~\cite{3GPPTR389012020}, i.e.  $\chi^\mathbf{x} \sim f^{\text{Gua}}\big(0,(\varsigma^{\mathbf{x}})^2\big)$.

%
%Computation model
%

\subsection{Computation Models}
\label{sec_DelayModel}

\par For task $\mathbf{\Psi}_v^t$ generated by vehicle $v$ at time slot $t$, it can either be processed by vehicle $v$ or offloaded to an MEC server based on the task offloading decision. To this end, we define the task offloading decision as a binary variable $k_{v,z}^t\in\{0,1\}, v\in\mathcal{V}, z\in\mathcal{M}\cup v$, where $\mathcal{M}\cup v$ represents the set of feasible destinations for vehicle $v$ to offload task $\mathbf{\Psi}_v^t$. In other words, $k_{v,v}^t=1$ indicates that the task is processed locally by vehicle $v$, while $k_{v,m}^t=1$ signifies that the task is offloaded to MEC server $m$ for remote processing. Moreover, both local computing and task offloading cause service delay and energy consumption, which can be elaborated as follows.

\subsubsection{Local Computing} 

\par The service delay and energy consumption for local computing are given as follows.

\par \textit{\textbf{Service Delay.}} When task $\mathbf{\Psi}_v^t$ is processed locally on vehicle $v$, the service delay primarily arises from local computing, which can be given as
\begin{equation}
	\label{eq_local_delay}
	\begin{aligned}
            D_{v,v}^t=l_v^t\mu_{v}^t/F_v.
		\end{aligned}
\end{equation}

\textbf{Energy Consumption.} According to~\cite{Sun2024Joint}, the energy consumption of vehicle $v$ to execute task $\mathbf{\Psi}_v^t$ can be given a
\begin{equation}
	\label{eq_energyLocalExe} 
        E_{v,v}^t=\gamma_v(F_v)^{2}l_v^t\mu_{v}^t,
\end{equation}

\noindent where $\gamma_i\geq0$ represents the effective capacitance of the CPU, which depends on the CPU chip architecture of vehicle $v$.

\subsubsection{Task Offloading} 

\par When task $\mathbf{\Psi}_v^t$ is offloaded to MEC server $m$, the MEC server is responsible for processing the task and returning the result. Note that we neglect the delay for result feedback, as the size of results for most mobile applications is significantly smaller than that of the input data.

\textbf{Service Delay.} When task $\mathbf{\Psi}_v^t$ is offloaded to MEC server $m$, the service delay mainly consists of the task offloading delay and task computation delay, which can be given as
\begin{equation}
	\label{eq_edge_delay}
	\begin{aligned}
            D_{v,m}^t=l_v^t/R_{v,m}^t+l_v^t\mu_{v}^t/f_{m,v}^t,
		\end{aligned}
\end{equation}

\noindent where $f_{m,v}^t$ denotes the computing resources allocated by MEC server $m$ to vehicle $v$ at time slot $t$.

\textbf{Energy Consumption.} The energy consumption for task offloading primarily consists of the transmission energy of vehicle $v$ and the computation energy of MEC server $m$, which can be calculated as            
\begin{equation}
	\label{eq_edge_energy} 
        E_{m,v}^t=p_v^tl_v^t/R_{v,m}^t+\epsilon l_v^t\mu_{v}^t,
\end{equation}

\noindent where $\epsilon$ represents the energy consumed per CPU cycle by MEC server $m$.

\subsection{Performance Metrics}

\par In this work, we construct the system cost by integrating service latency and energy consumption to evaluate the performance of the proposed system. Specifically, the service delay for completing task $\mathbf{\Psi}_v^t$ can be calculated based on Eqs.~\eqref{eq_local_delay} and \eqref{eq_edge_delay} as follows:
\begin{equation}
	\label{eq_service_delay}
	\begin{aligned}
D_v^t=k_{v,v}^tl_v^t\mu_{v}^t/F_v+\sum_{m\in\mathcal{M}}k_{v,m}^t\big(l_v^t/R_{v,m}^t+l_v^t\mu_{v}^t/f_{m,v}^t\big).
		\end{aligned}
\end{equation}

\noindent Moreover, the energy consumption for completing task $\mathbf{\Psi}_v^t$ can be calculated based on Eqs.~\eqref{eq_energyLocalExe} and \eqref{eq_edge_energy} as follows:
\begin{equation}
	\label{eq_energy_consumption}
	\begin{aligned}
E_v^t=k_{v,v}^t\gamma_v(F_v)^{2}l_v^t\mu_{v}^t+\sum_{s\in\mathcal{S}}o_{v,s}^t(p_v^tl_v^t/R_{v,m}^t+\epsilon l_v^t\mu_{v}^t).
		\end{aligned}
\end{equation}

\par Similarly to~\cite{he2024qoe}, the system cost of completing task $\mathbf{\Psi}_v^t$ at time slot $t$ can be presented as
\begin{equation}
	\label{eq_cost}
	\begin{aligned} 
         C_v^t=w_v^{\text{D}}D_v^t+w_v^{\text{E}}E_v^t,
		\end{aligned}
\end{equation}

\noindent where $w_v^{\text{D}}$ and $w_v^{\text{E}}$ respectively represent the weights of service delay and energy consumption, which can be adjusted according to the preference of the system regarding the trade-off between delay and energy consumption costs.

\section{Problem Formulation and Analysis}
\label{sec_problem_analysis}

\par In this section, we present the problem formulation and analysis.

%
% Problem formulation
%
\subsection{Problem Formulation}
\label{sec_problemFormulation}

\par The optimization problem is formulated to minimize the time-averaged system cost by jointly determining the strategies of task offloading  $\mathbf{K}= \{k_{v,z}^t\}_{v\in \mathcal{V}, z \in \mathcal{M}\cup v, t \in \mathcal{T}}$ and computing resource allocation $\mathbf{F}= \{f_{m,v}^t\}_{m\in \mathcal{M}, v\in \mathcal{V}, t \in \mathcal{T}}$ over $T$ time slots. Accordingly, the problem can be formulated as follows:
\begin{subequations}
	\label{eq_problem}
	\begin{alignat}{2}
		\mathbf{P}: \quad &\min_{\mathbf{K},\mathbf{F}}  \frac{1}{T}\sum_{t=1}^T\sum_{v=1}^V C_v^t, \label{utility}\\
  \text{s.t.} \quad &  k_{v,z}^t\in\{0,1\}, \ \forall v\in \mathcal{V}, \, z\in \mathcal{M}\cup v, \, t\in \mathcal{T},\label{eq_c1}\\
		& \sum_{v\in \mathcal{V}}k_{v,m}^t\leq 1, \, \forall  m\in \mathcal{M},  \, t\in \mathcal{T}, \label{eq_c2}\\
		& D_{v}^t\leq \tau_v^t, \, \forall v\in \mathcal{V},  \, t\in \mathcal{T},  \label{pro_c4}\\
		&\sum_{v\in\mathcal{V}} k_{v,m}^tf_{v,m}^t \leq  F_m, \, \forall m \in \mathcal{M},  \, t\in \mathcal{T}, \label{pro_c6}\\
		&\sum_{v \in \mathcal{V}} k_{v,m}^t E_{m,v}^t\leq E_m, \, \forall m\in \mathcal{M},  \, t\in \mathcal{T}, \label{pro_c8}\\ 
      &\sum_{v \in \mathcal{V}} k_{v,v}^t E_{v}^t\leq E_v, \, \forall v\in \mathcal{V},  \, t\in \mathcal{T}, \label{pro_c9}
  \end{alignat}
\end{subequations}

\noindent where constraints~\eqref{eq_c1} and ~\eqref{eq_c2} indicate that the decision of task offloading is binary. Moreover, constraint~\eqref{pro_c4} ensures that the service delay should not exceed the deadline of task. In addition, constraints~\eqref{pro_c6}imposes limits on the computing resources for the MEC server. Besides, \eqref{pro_c8} and \eqref{pro_c9} denote the energy constraints of each vehicle and MEC server, respectively.

\subsection{Problem Analysis}

\par Solving problem $\mathbf{P}$ directly may presents several challenges as follows: 

\begin{itemize}
\item \textit{NP-hard and non-convex MINLP.} Problem $\mathbf{P}$ involves both binary decision variables (i.e., task offloading $\mathbf{K}$) and continuous decision variables (i.e., computing resource allocation $\mathbf{F}$) is an MINLP problem, which is an NP-hard and non-convex~\cite{he2024space}. This makes it challenging to find an optimal solution of this problem. 

\item \textit{Long-term objective in a highly dynamic environment.} The optimization problem aims to minimize the long-term system cost in the MEC-assisted consumer IoV system, which is characterized by highly dynamics such as time-varying wireless channels, real-time vehicle mobility, and random task arrivals. This dynamic leads to partial observability of the environment, posing a significant challenge in designing a real-time approach that can dynamically adapt to unknown and changing environment. 

\item \textit{Mutual coupling of decision variables across multiple nodes.} In problem $\textbf{P}$, the decision variables of different vehicles are interdependent and mutually coupled. On the one hand, the offloading decisions of vehicles and the computing resource allocation decisions of MEC servers collectively influence the overall system cost. On the other hand, the decisions of different nodes have mutual impact on each other. 

%For example, the decision of computing resource allocation affects service delay, which in turn influences offloading decisions of vehicles. Additionally, for a certain vehicle, the offloading decisions of its neighboring vehicles can impact the computing resource allocation of MEC servers, which further impacts the offloading decision of the vehicle. This mutual coupling decisions across multiple nodes adds complexity to efficient decision-making process.
\end{itemize}

%
%Algorithm
%
\section{The Proposed JTOCRA}
\label{sec_jointOffloading}

\par In this section, we reformulate the optimization problem in \eqref{eq_problem} as a universal Markov game and propose an MADDPG algorithm to obtain the solutions of the problem.

\subsubsection{Motivation}

\par The proposed JTOCRA is guided by the following key motivations:

\begin{itemize}
\item \textit{Challenges of NP-hard and non-convex MINLP in $\mathbf{P}$.} The NP-hard and non-convex MINLP feature of problem $\mathbf{P}$ poses significant challenges in obtaining an optimal solution. Although conventional theoretical optimization algorithms, such as convex optimization, have been widely used to address these problems, they present several critical limitations. Specifically, the conventional optimization approaches rely heavily on theoretical models, which struggle to capture the real-time characteristics for dynamic environments. Moreover, even though more sophisticated models can be constructed to better capture the complexity features of the dynamic scenarios, problem decomposition is typically required to decrease the computation complexity, which decreases the solution space and often leads to sub-optimal outcome.

\item \textit{Leveraging DRL for the long-term objective in dynamic environments.} The goal of optimizing long-term system costs in a highly dynamic MEC-assisted consumer IoV system motivates the adoption of DRL-based approaches. Specifically, DRL is inherently designed to optimize cumulative rewards over time, making it well-suited for achieving long-term objectives. Additionally, DRL is able to effectively  adapt to the dynamic scenarios by learning strategies through continuous interaction with the environment. Furthermore, the techniques such as experience replay, efficient neural network architectures, and model-free learning enable DRL to make real-time decisions, effectively addressing the delay-sensitive requirements of MEC-assisted IoV systems.

\item \textit{MADDPG for mutual-coupled decision variables in the multi-user IoV system.} The mutual coupling of decision variables across multiple nodes in the MEC-assisted IoV system further motivates the use of MADDPG. Specifically,  MADDPG effectively captures the interactions and dependencies among different nodes. By this way, each node are able to make decisions based on the actions of the others, resulting in more sophisticated and stable outcomes. Moreover, MADDPG allows each node to operate in a decentralized manner based on its local observations, making the approach more adaptive to the dynamic IoV environment where full knowledge of the global state is not always available. Consequently, MADDPG is well-suited to handle the complexity arising from the coupling of decisions across multiple nodes and to adapt to the dynamic nature of the system, making it suitable for the dynamic and multi-user MEC-assisted IoV system.

\end{itemize}

\subsection{Problem Reformulation}
\label{sec_MDP}
\par We model the problem of minimizing the long-term total cost in vehicular networks as a Markov game, which is an extension of the Markov decision process (MDP). Specifically, the Markov Game model can be described by a tuple $\langle \mathcal{S}, \mathcal{N}, \mathcal{O}, \mathcal{A}, \mathcal{P}, \mathcal{R}, \gamma \rangle$, where $\mathcal{S}$ represents the set of all possible states of the environment, $\mathcal{N}$ denotes the set of agents (each vehicle is considered as an individual agent in this work), $\mathcal{O}$ denotes the set of observations available to each agent, $\mathcal{A}$ means the actions available to the agents, $\mathcal{P}$ indicates the state transition probability function, $\mathcal{R}$ represents the reward function, and $\gamma$ is the reward discount factor.

\par At time slot $t$, each agent $n \in \mathcal{N}$ observes the environment through the observation function $o_n$, defined as ${o^t}_n \triangleq f(s^t,n)$, where $ s^t \in \mathcal{S} $ denotes the environment state in  time slot $t$. Based on this observation, agent $n$ selects an action $a^t_n$ according to its policy $\pi_n(o_n^t)$. After that, the joint action $a^t$ based on the actions selected by all agents is then executed, leading the environment to transition to the next state $s^{t+1}$ according to the transition probability function $ P(s^{t+1} \mid s^t, a^t)$. Subsequently, the agent  receives a reward $r_n^t$ from the environment based on its action and the new state. The key components of the Markov game model are detailed below.

\subsubsection{Agent} In the MEC-assisted consumer IoV system, each vehicle is regarded as an agent with its own actor network to determine the actions, while all agents share the critic network for policy evaluation.

\subsubsection{State} The global environment state consists of the information about the vehicles (including position $\mathbf{q}_v^t$, computing resource constraint $F_v$, energy constraint $E_v$), the task $\mathbf{\Psi}_v^t$ generated by each vehicle at the current time (including data size $l_v^t$, computational density $\mu_v^t$, and the deadline of the task $\tau_v^t$), and the MEC servers (including position $\mathbf{q}_m$, computing resources $F_m$, energy constraint $E_m$). Therefore, the global environment state can be represented as 
\begin{sequation}
\begin{aligned}
    s^t=&\{\mathbf{q}_v^t,F_v,E_v,l_v^t,\mu_v^t,\tau_v^t, \mathbf{q}_m,F_m,E_m\}, \\ &\forall s\in \mathcal{S}, \forall v \in \mathcal{V}, \forall m \in \mathcal{M}.
\end{aligned}
\end{sequation}

\subsubsection{Observation} For each agent, only partial state information can be observed from the global state of the system. Specifically, the observation space for each agent $n$ includes its position, its available computing resources, the state of the generated task, and the position of the associated MEC server, which can be represented as
\begin{equation}
o_n^t=\{\mathbf{q}_n^t, l_n^t, \mu_n^t, \tau_n^t, F_n, \textbf{q}_m^t\}, \forall n \in \mathcal{N}, \forall m \in \mathcal{M}.
\end{equation}

\subsubsection{Action} At time slot $t$, each vehicle needs to make decisions regarding task offloading and the requested computing resource. Thus, the action space for each agent $n$ can be represented as 
\begin{equation}
\begin{aligned}
a_n^t=\{k_n^t,f_n^t\}, \forall n\in \mathcal{N}.
\end{aligned}
\end{equation}
\subsubsection{Reward} Each agent receives its own reward from the environment at each time slot. To stimulate cooperation among the agents, we consider that all agents receive the same reward $r^t$ based on a reward function, i.e., $r_1^t=r_2^t=\dots=r_n^t$. According to the objective function of system cost minimization for problem $\mathbf{P}$, the reward of agents consists of both extrinsic reward and penalty, which can be represented as
\begin{equation}
	\label{eq_reward} 
        r^t = r_E^t + r_P^t,
\end{equation}

\noindent where $r_E^t $ and $ r_P^t$ denote the extrinsic reward and penalty, respectively. Moreover, the detailed model for $r_E^t $ and $ r_P^t$ are detailed as follows.

\begin{itemize}
    \item \textit{Extrinsic reward:} Extrinsic rewards are typically related to the objective function. For the purpose of minimizing the system cost, the extrinsic reward can be defined as 
    \begin{equation}
        \label{eq_extrinsic_rewawrd}
        r_E^t = \sum_{v=1}^\mathcal{V}C_v^t.
    \end{equation}
    
    \item \textit{Penalty}: 
    In the considered system, the penalties include those imposed for task timeout and for exceeding energy constraints of vehicles and MEC servers, which can be given as follows:
    \begin{equation}
        \label{eq_penalty}
    r_P^t=\sum_{v=1}^\mathcal{V}\left(\phi_TD_v^{t,out}+\phi_EE_v^{t,out}\right) + \sum_{s\in \mathcal{S}}\phi_EE_s^{t,out},
    \end{equation}
    
    \noindent where $\phi_T$ and $\phi_E$ denote the penalty coefficients for exceeding the limits of task deadlines and energy consumption constraints, respectively. Additionally, $D_v^{t,out}$ represents the portion of task $\mathbf{\Psi}_v^t$ that exceeds the deadline. Besides, $E_v^{t,out}$ and $E_s^{t,out}$ correspond to the energy consumption exceeding the constraints for vehicles and MEC servers, respectively.
\end{itemize}

\subsubsection{Transition Probability}

\par In the considered Markov game model, the transition probability of a state is defined as $P(s^{t+1}|s^t, a_1^t, \dots, a_n^t)$, which represents the probability distribution of the next state $s^{t+1}$ after all agents have performed their actions in the current state $s^t$.

\subsection{MADDPG Algorithm}
\label{sec_MADDPG}

\par  In this section, MADDPG is employed to solve the above reformulated problem since it can effectively deal with the NP-hard and non-convex MINLP, optimizes long-term objective of system cost minimization in dynamic IoV environment, and manages the mutual coupling of decision variables across multiple vehicles, as mentioned in Section \ref{sec_problem_analysis}. MADDPG integrates three key components, i.e., off-policy learning, actor-critic architecture, and centralized training with decentralized execution (CTDE).  Specifically, MADDPG employs a centralized critic network during training, which has access to the global state and the actions of all agents. This enables critic to fully consider the interactions and coordination among agents to make efficient decisions. In contrast, the actor network of each agent only uses local observation information, enabling agents to make independent decisions in real time based on their own observations.

%\par Multi-Agent Deep Deterministic Policy Gradient (MADDPG) is a deep reinforcement learning algorithm specifically designed for multi-agent environments, building upon the DDPG framework. The algorithm integrates three key concepts: Off-Policy Learning, Actor-Critic Architecture, and Centralized Training with Decentralized Execution (CTDE). MADDPG aims to address the unique challenges in multi-agent systems, particularly the instability that arises when agents treat other agents as part of the environment, leading to an inability to account for the dependencies and interactions between agents.

%\par To mitigate these issues, MADDPG employs a centralized critic network during training, which has access to the global state and the actions of all agents. This allows the critic to fully consider the interactions and coordination required among agents. In contrast, each agent's actor network only uses local observation information, enabling agents to make independent decisions based on their own observations during execution, without needing information about other agent $s^{\prime}$ strategies. This approach not only enhances system stability but also improves learning efficiency and performance in complex multi-agent cooperative tasks.

\subsection{Network Architecture of MADDPG}

\par The network architecture of each agent $n\in\mathcal{N}$ is structured as below.

\subsubsection {Critic Network} 

\par The critic network takes the joint observations $o^t = \{o_1^t, \dots, o_n^t\}$ and the joint actions $a^t = \{a_1^t, \dots, a_n^t\}$ from all agents. To enhance training stability and mitigate the Q-value overestimation issue, the critic network of each agent is composed of two sub-networks, i.e., an online critic network $Q_{\theta_n}$ with parameters $\theta_n$, and a corresponding target critic network $Q^\prime_{\theta^\prime_n}$ with parameters $\theta^\prime_n$.

\subsubsection{Actor Network} The actor network for each agent takes its own local observation $o_n^t$ as inputs and outputs of the action $a_n^t$. Similar to the critic network, the actor network also includes an online actor network $\mu_{\phi_n}$ with parameters $\phi_n$ and a target actor network $\mu^\prime_{\phi^\prime_n}$ with parameters $\phi^\prime_n$. The target networks are used to stabilize the training by gradually  tracking the updates of the online networks.

\subsection{Training Process of MADDPG}

\par During the training phase, agents utilize additional information, such as the observations and actions of the other agents, to coordinate their action selections. Once all agents have made the decisions, they receive corresponding rewards, which results in the environment transitioning to the next state $s^{t+1}$. Then, the experience tuple $(s^t, s^{t+1}, a^t, r^t)$ is recorded in the replay buffer $\mathcal{D}$. Subsequently, agents randomly sample a batch $\mathcal{B}$ of size $|B|$ from the replay buffer $\mathcal{D}$ to update their network parameters. During the execution phase, agents choose the optimal actions directly based on their own observations and the learned policy.

\subsubsection{Critic Network Update}

\par The online critic network $Q_{\theta_{n}}$ is updated based on the temporal difference error, which quantifies the discrepancy between the predicted Q-value and the target value. Specifically, the temporal difference error is calculated by using the loss function for the critic network, which is defined as
\begin{equation}
    \label{eq_critic_loss}
    L(\theta_{n}) = \mathbb{E} [(Y_n^t - Q_{\theta_{n}}(s^t, a^t))^2],
\end{equation}

\noindent where $Y_n^t$ represents the target value, which is calculated using the target critic network and target actor network as
\begin{equation}
    \label{eq_target_value}
    Y_n^t = r^t + \gamma Q^\prime_{\theta^\prime_{n}} \big( s^{t+1}, \mu^\prime_{\phi^\prime_{n}}(o_n^t) \big),
\end{equation}

\noindent where $\gamma$ represents the discount factor, $Q^\prime_{\theta^\prime_{n}}$ denotes the target critic network, and $\mu^\prime_{\phi^\prime_{n}}$ is the target actor network.

\par The gradient of the loss function with respect to the critic parameters $\theta_{n}$ can be given as
\begin{sequation}
\begin{aligned}
    \label{eq_gradient_critic}
    &\nabla_{\theta_{n}} L(\theta_{n}) \\&= -2 \mathbb{E} [r^t + \gamma Q^\prime_{\theta^\prime_{n}}(s^{t+1},a^t) - Q_{\theta_{n}}(s^t, a^t) \nabla_{\theta_{n}} Q_{\theta_{n}}(s^t, a^t)].
\end{aligned}
\end{sequation}

\par Accordingly, the critic network parameters is updated as
\begin{equation}
    \label{eq_critic_update}
    \theta_{n} \leftarrow \theta_{n} - \xi_c \nabla_{\theta_{n}} L
\end{equation}

\noindent where $\xi_c$ denotes the learning rate for the critic network. 

\subsubsection{Actor Network Update}

\par The objective function for the actor network is to maximize the expected Q-value of the actions chosen by the policy, which can be expressed as
\begin{equation}
    \label{eq_actor_function}
    \pi_n^*=\arg\max_{\phi_n} J_{\pi_n}(\phi_n),
\end{equation}

\noindent where $\pi_n$ represents the action policy for agent $n$, and $J_{\pi_n}(\phi_n)$ denotes the policy objective function with respect to the actor parameters $\theta_{\mu_j}$. The gradient of this objective function can be calculated as
\begin{sequation}
    \label{eq_actor_gradient}
    \nabla_{\phi{n}} J_{\pi_n}(\phi_n) = \mathbb{E} [ \nabla_{a_n^t} Q_{\theta_{n}}(s^t, a^t) |_{a_n^t = \mu_{\phi_{n}}(o_n^t)} \nabla_{\phi_{n}} \mu_{\phi_{n}}(o_n^t) ].
\end{sequation}

\par Accordingly, the actor parameters $\phi_{n}$ are updated as
\begin{equation}
    \label{eq_actor_update}
    \phi_{n} \leftarrow \phi_{n} + \xi_a \nabla_{\phi_{n}} J(\phi_{n}),
\end{equation}

\noindent where $\xi_a$ represents the learning rate for the actor network. 

\subsubsection{Stabilization with Target Networks}

\par To ensure stability during training, the target networks are updated gradually to track the online networks. Specifically, the update rule of the critic parameter $\theta_{n}$ for the target networks are as follows:
\begin{equation}
    \label{eq_critic_target}
    \theta^\prime_{n} \leftarrow \lambda \theta_{n} + (1 - \lambda) \theta^\prime_{n},
\end{equation}

\noindent where $\lambda$ is the update rate for the target networks. 

\par Moreover, the update rule of the actor parameter $\theta_{n}$ for the target networks are as follows:

\begin{equation}
    \label{eq_actor_target}
    \phi^\prime_{n}\leftarrow \lambda \phi_{n} + (1 - \lambda) \phi^\prime_{n}.
\end{equation}

\subsection{Main Steps of JTOCRA and Complexity Analysis}
\label{sec_complexity}

\par
\begin{figure}[!hbt] 
    \centering
    \setlength{\abovecaptionskip}{0pt}%    
    \setlength{\belowcaptionskip}{0pt}%
    \includegraphics[width =3.5 in]{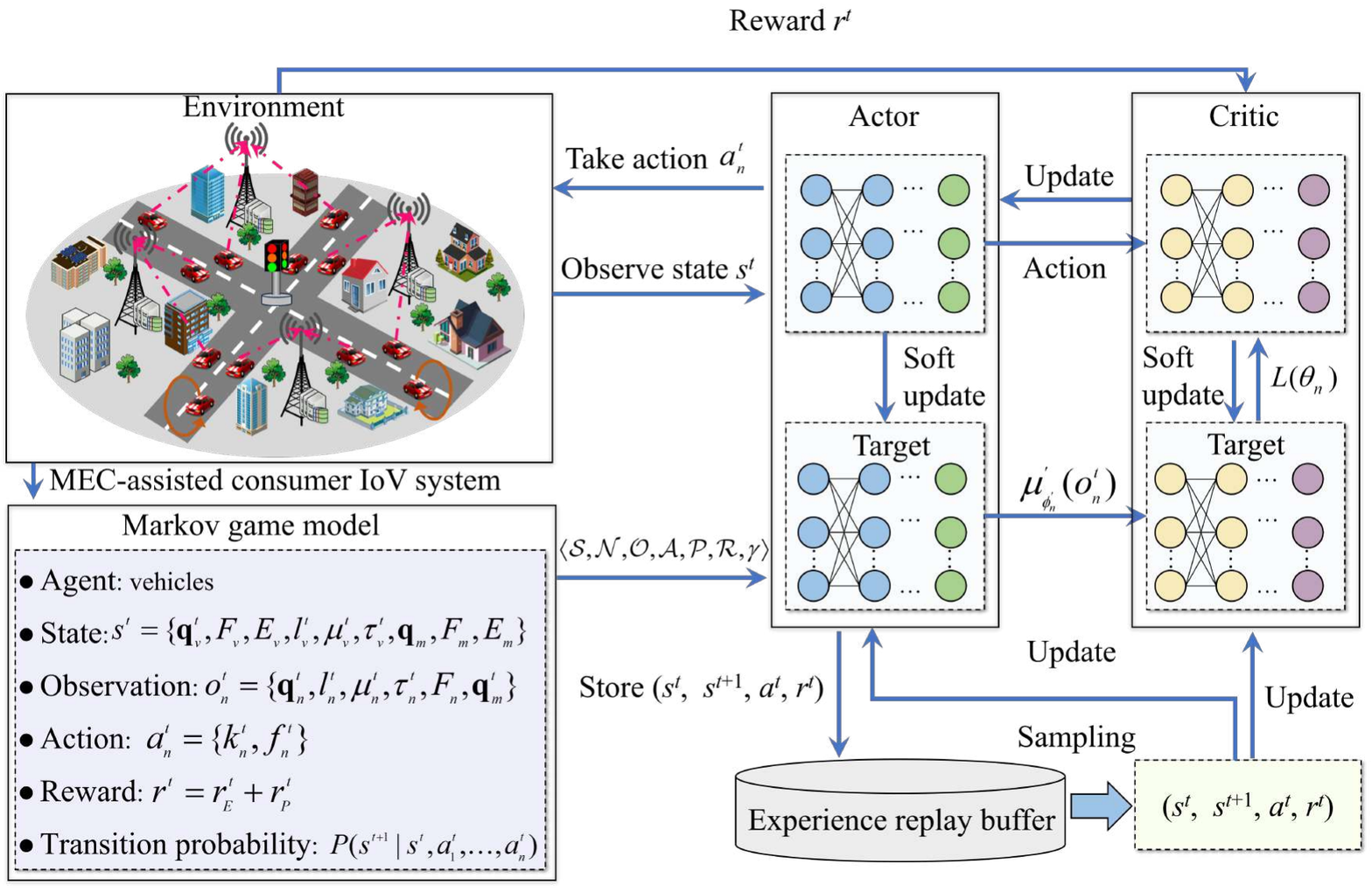}
    \caption{\textcolor{b}{Framework of JTOCRA in the MEC-assisted consumer IoV system.}}
    \label{fig_framework}
    % \vspace{-1.5em}
\end{figure}

 Fig. \ref{fig_framework} shows the framework and main flow of the proposed JTOCRA, and the main steps of JTOCRA \textcolor{b}{are} described in Algorithm \ref{JTOCRA}. Moreover, the computational complexity analysis of the proposed JTOCRA approach includes two primary components, i.e., the training phase and the execution phase. First, during the training phase, each agent updates both its actor and critic networks based on experiences sampled from the replay buffer. This process involves forward and backward propagation through these networks, gradient computation, and parameter updates. Therefore, the computational complexity for training the actor and critics networks for one agent during one training iteration can be given by $ O (M T B L_A  N_A + L_C  N_C t))$,  where $L_A$ and $N_A$ denote the number of layers and neurons per layer in the actor network, respectively. Moreover, $L_C$ and $N_C$ represent the number of layers and neurons per layer in the critic network, respectively. Additionally, $M$ indicates the number of episodes, $T$ means the length of each episode, and $B$ is the batch size. Second, during the execution phase, each agent uses its actor network to select actions based on the current observation. Therefore, the computational complexity for generating actions during execution for an agent can be expressed as $O(L_A  N_A)$. 

%Specifically, for each vehicle, the critic network, actor network, and target network are initialized (lines 2 to 4). Then, in each episode, each vehicle makes independent decisions based on the observations and learned actions (lines 5 to 11). Furthermore, the reward of the next state \textcolor{b}{is} received and the experience tuple is stored. \textcolor{b}{Besides, the critic network}, actor network, and the target networks are updated (lines 14 to 18).

% \subsubsection{Computational Complexity of JTOCRA}

% \par The computational complexity analysis of the proposed JTOCRA approach includes two primary components, i.e., the training phase and the execution phase. First, during the training phase, each agent updates both its actor and critic networks based on experiences sampled from the replay buffer. This process involves forward and backward propagation through these networks, gradient computation, and parameter updates. Therefore, the computational complexity for training the actor and critics networks for one agent during one training iteration can be given by $ O\left( M T  B  \left( L_A  N_A + L_C  N_C \right) \right)$,  where $L_A$ and $N_A$ denote the number of layers and neurons per layer in the actor network, respectively. Moreover, $L_C$ and $N_C$ represent the number of layers and neurons per layer in the critic network, respectively. Additionally, $M$ indicates the number of episodes, $T$ means the length of each episode, and $B$ is the batch size. Second, during the execution phase, each agent uses its actor network to select actions based on the current observation. This process involves only the forward pass through the actor network, making the execution phase less computationally intensive compared to the training phase. Therefore, the computational complexity for generating actions during execution for an agent can be expressed as $O\left( L_A  N_A \right)$. 

\begin{algorithm}[]	
	\label{JTOCRA}	
	\SetAlgoLined
        Initialize experience replay buffer $\mathcal{D}$ with size $C$.\\
	\For{$n \in \mathcal{N}$}
	{
        Initialize online critic network $Q(\theta_n)$ and actor network $\mu(\phi_n)$ with parameters $\theta_n$ and $\phi_n$ for agent $n$. \\
        Initialize the corresponding target networks $Q(\theta^\prime_n)$ and $\mu^\prime(\phi_n)$ by $\theta^\prime_n \leftarrow \theta_n$ and $\phi^\prime_n \leftarrow \phi_n$.
	}
	\For{each episode}
	{
        Reset the initial state $s^0$ of the system. \\
        \While{$t \leq T$}
        {
            \For{$n \in \mathcal{N}$}
            {
                Get the observation $o_n^t$. \\
                Select action $a_n^t = \pi_n(o_n^t) + \eta^t$ with noise $\eta^t$.
            }
            Execute joint action $a^t = (a_1^t, \cdots, a_n^t)$. \\
            Receive immediate reward $r_t$ and next state $s^{t+1}$. \\
            Store the experience tuple $(s^t, s^{t+1}, a^t, r^t)$ to the replay buffer $\mathcal{D}$. \\
            \For{$n \in \mathcal{N}$}
            {
                Sample a random batch of size $|B|$ from the replay buffer $\mathcal{D}$. \\
                Update the online critic network by (\ref{eq_critic_update}). \\
                Update the online actor network by (\ref{eq_actor_update}). \\
                Update the corresponding target networks by (\ref{eq_critic_target}) and (\ref{eq_actor_target}).
            }
        }
    }    
	\caption{\textcolor{b}{JTOCRA}.}
\end{algorithm}

\section{Simulation Results and Analysis}
\label{sec_simulation}

% \par In this section, we conduct simulation experiments to validate the effectiveness of the proposed approach.

\subsection{Simulation Setup}
\label{simulation_set_up}

\par We first present the settings of the simulation experiments, including the scenario, parameters, comparative approaches, and performance evaluation metrics.

\subsubsection{Scenarios} 

\par We consider an MEC-assisted consumer IoV system, where $4$ MEC servers are deployed along side the road to provide offloading service for $20$ moving vehicles within a $1000\times1000$ square area. 

\subsubsection{Parameters} 
% \vspace{-2em}
\par For the MEC servers, the horizontal positions are set as [500, 500], [500, 500], [500, 500], and [500, 500], respectively. For vehicles, the initial positions are set randomly in the considered area, and the moving velocities are initialized in the range of [10, 25] m/s. The default values of the other parameters are listed in Table \ref{tab_parameter}.

\begin{table}[!hbp]
	\setlength{\abovecaptionskip}{0pt}%    
	\setlength{\belowcaptionskip}{0pt}%
	\caption{Simulation parameters}
	\label{tab_parameter}
	\renewcommand*{\arraystretch}{1}
	\begin{center}
		\begin{tabular}{p{.07\textwidth}|p{.22\textwidth}|p{.12\textwidth}}
			\hline
			\hline
			\textbf{Symbol}&\textbf{Description}&\textbf{Default value}\\
               \hline
                $F_v$&Computing resources of vehicle $v$& $[1,5]$ GHz\\
                \hline
				 $E_v$&Energy constraint of vehicle $v$ & $[5, 25]$ J\\
			\hline
			  	$l_v^t$&Task size & [1, 3] Mb \\
			  \hline
                 $\mu_{v}^t$&Computation intensity of tasks&  [500, 1500] cycles/bit \\ 
			  \hline
                  $\tau_{v}^t$&Deadline of task&  [1,5] s \\ 
			  \hline	 
				  $E_m$& Energy constraint of MEC server $m$&1000 J \\
			  \hline
				  $F_{m}$&Computing resources of MEC server $m$&[50, 100] GHz \\
                \hline
                    $B_{v,m}$&Bandwidth between vehicle $v$ and MEC server $m$ & 20 MHz  \\
                \hline
                 $P_{v}$ &Transmit power of vehicle $v$ & [10,25] dBm\\
                 \hline
			         $N_0$&Noise power& -98 dBm\\
			\hline
   				$\alpha_1$/$\alpha_2$&Environment parameters&18 m/36 m \cite{3GPPTR1389012018}\\
                \hline   
				    $\tau_i^t$&The deadline of task &[0.1, 5] s\\
			  \hline
			         $\omega$ &Memory degree of vehicle velocity&0.9 \\
			   \hline
			  	 $\sigma$ & Standard \textcolor{b}{deviation} of velocity&2\\
			   \hline
				    $\gamma_v$&Effective capacitance of the CPU for vehicle $v$& $10^{-28}$\\
			\hline
                 $\epsilon$&The energy consumed per CPU cycle by MEC servers&$8.2\times10^{-28}$ J\\
                \hline
                 $\xi_c$&Learning rate of critic network&$5\times10^{-4}$ \\
                \hline
                 $\xi_a$&Learning rate of actor network&$5\times10^{-4}$ \\
                \hline
                 $\eta$&Soft update rate&$5\times10^{-3}$ 
                 \\
                \hline
                 $C$&Size of replay buffer&$10^{5}$\\
                \hline
                 $|B|$&Batch size&128\\
                 \hline
                 \textcolor{b}{$\mathbf{m}^{\text{L}}/\mathbf{m}^{\text{N}}$}&\textcolor{b}{ Shaping factor of the Nakagami fading channel} &\textcolor{b}{$4/2$~\cite{peng2022directional}}\\
                \hline
				\textcolor{b}{$\beta^{\text{L}}/\beta^{\text{N}}$}&\textcolor{b}{Path loss exponent} &\textcolor{b}{$2.42/4.28$~\cite{yang2018dense}}\\
			\hline
                    \textcolor{b}{ $\varsigma^{\text{L}}/\varsigma^{\text{N}}$}&\textcolor{b}{Standard deviation of shadowing for LoS/NLoS communication}& \textcolor{b}{$4/6$~\cite{3GPPTR389012020}}\\	
                \hline
                 \textcolor{b}{$\omega$} & \textcolor{b}{Memory degree}& \textcolor{b}{0.9}\\
                 \hline
                 \textcolor{b}{$c$}& \textcolor{b}{Light speed} & \textcolor{b}{$3\times 10^8$ m/s}\\
                 \hline
                  \textcolor{b}{$\bar{d}$}&\textcolor{b}{Reference distance}&\textcolor{b}{1 m}\\ 
                \hline
                 \textcolor{b}{$f_c$}&\textcolor{b}{Carrier frequency}&\textcolor{b}{2 GHz}\\
                 \hline
		\end{tabular}
	\end{center}
\end{table} 

\subsubsection{Baselines} 

\par To validate the effectiveness of the proposed JTOCRA, we compare the JTOCRA with the following approaches.

\begin{itemize}
\item \textit{Nearest task offloading (NTO)}: The tasks of each vehicle are offloaded to the nearest MEC server, while the decision of computing resource allocation is decided by the proposed JTOCRA.

\item \textit{Equal computing resource allocation (ECRA)}: The computing resource of each MEC server are equally allocated, while the decision of task offloading is determined by the proposed JTOCRA.

\item \textit{MAPPO-based task offloading and computing resource allocation (MTOCRA)}: The task offloading and computing resource allocation are jointly decided by the multi-agent PPO (MAPPO) algorithm.

% 智能体不共享critic网络，

\item \textit{Independent DDPG-based task offloading and computing resource allocation (IDTOCRA)}: The task offloading and computing resource allocation are jointly decided by the distributed DDPG algorithm.

\end{itemize}

\subsubsection{Performance Indicators} 

\par We evaluate the performance of the proposed JTOCRA by considering the following metric. \textit{i) Average system cost} $\frac{1}{T}\sum_{t=1}^T\sum_{v=1}^V C_v^t$, which is the objective function and measures the overall system performance. \textit{ii) Average service delay} $\frac{1}{T}\sum_{t=1}^T\sum_{v=1}^V D_v^t$, which evaluates the service delay for task offloading. \textit{iii) Average energy consumption} $\frac{1}{T}\sum_{t=1}^T\sum_{v=1}^V E_v^t$, which evaluates the total energy consumption of both vehicles and MEC servers. 

\subsection{Performance Evaluation}

\par In this section, we first evaluate the system performance of the proposed JTOCRA over time with default parameters. Subsequently, we compare the impacts of different parameters on the performance of the proposed JTOCRA and the benchmark approaches.

\subsubsection{Convergence and Stability Verification}
 
\par
\begin{figure}[!hbt] 
    \centering
    \setlength{\abovecaptionskip}{0pt}%    
    \setlength{\belowcaptionskip}{0pt}%
    \includegraphics[width =3.1 in]{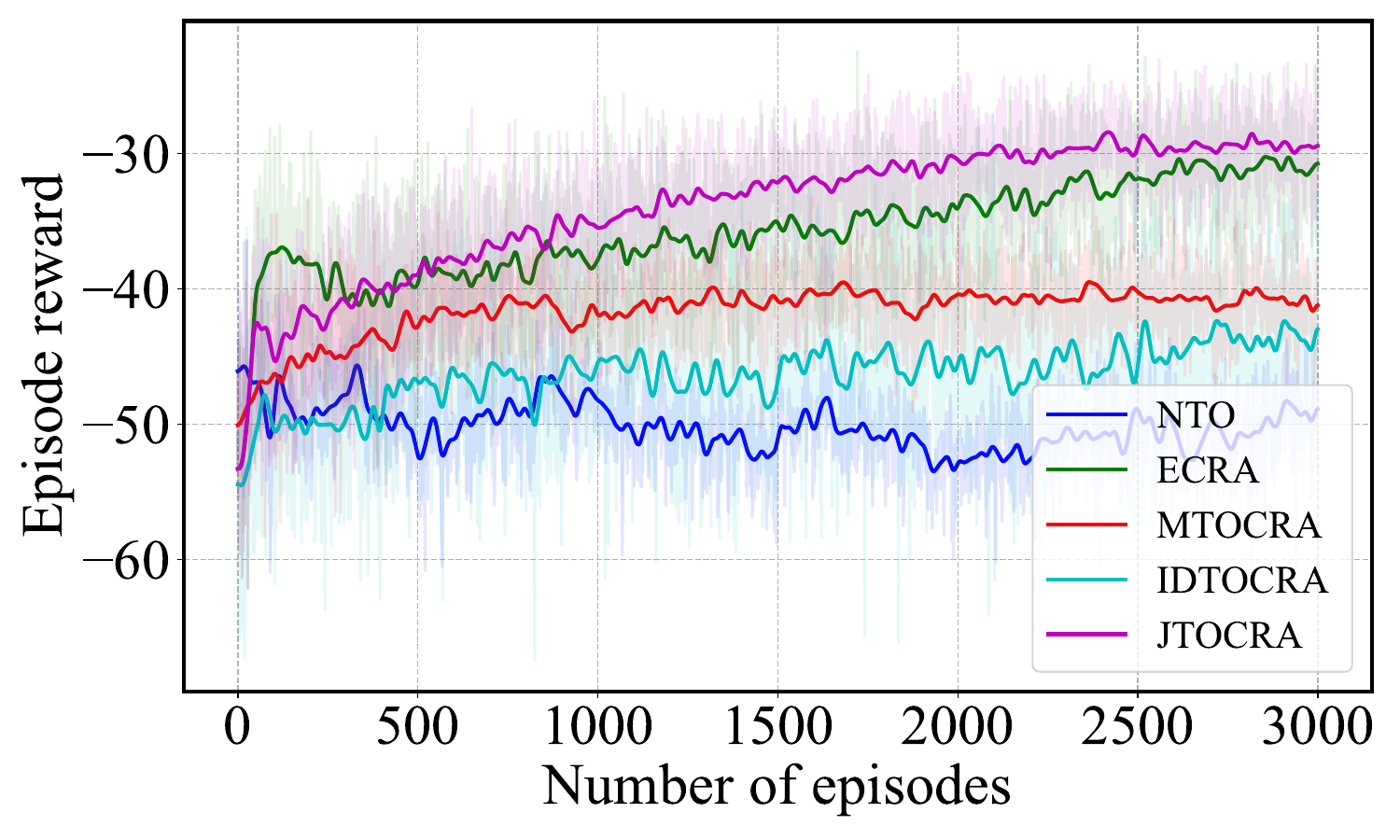}
    \caption{Convergence performance.}
    \label{fig_converge}
    \vspace{-0.5em}
\end{figure}

 Fig.~\ref{fig_converge} illustrates the convergence for the five approaches. It can be observed from the figure that all five approaches converge at approximately 300 epochs. However, the proposed JTOCRA achieves the highest reward value, achieving approximately 42.20\%, 5.52\%, 27.87\%, and 32.91\% performance gains compared with NTO, ECRA, MTOCRA, and IDTOCRA, respectively at 3000 episodes. This superior performance is primarily due to the ability of the JTOCRA to jointly consider the coupled decisions of multiple vehicles while enabling distributed action execution, making it particularly well-suited for the dynamic IoV scenario. In summary, the proposed JTOCRA demonstrates superior convergence and stability performance among the five approaches with a superior reward value.

\subsubsection{Impact of Parameters}

\par In this section, we compare impacts of task data size, MEC server computing resource, and the number of vehicles on the performance of the proposed JTOCRA and the benchmark approaches.

\begin{figure*}[!hbt] 
	\centering
	\subfigure[Average system cost]
	{
		\begin{minipage}[t]{0.3\linewidth}
			\raggedleft
			\includegraphics[width=2.2in]{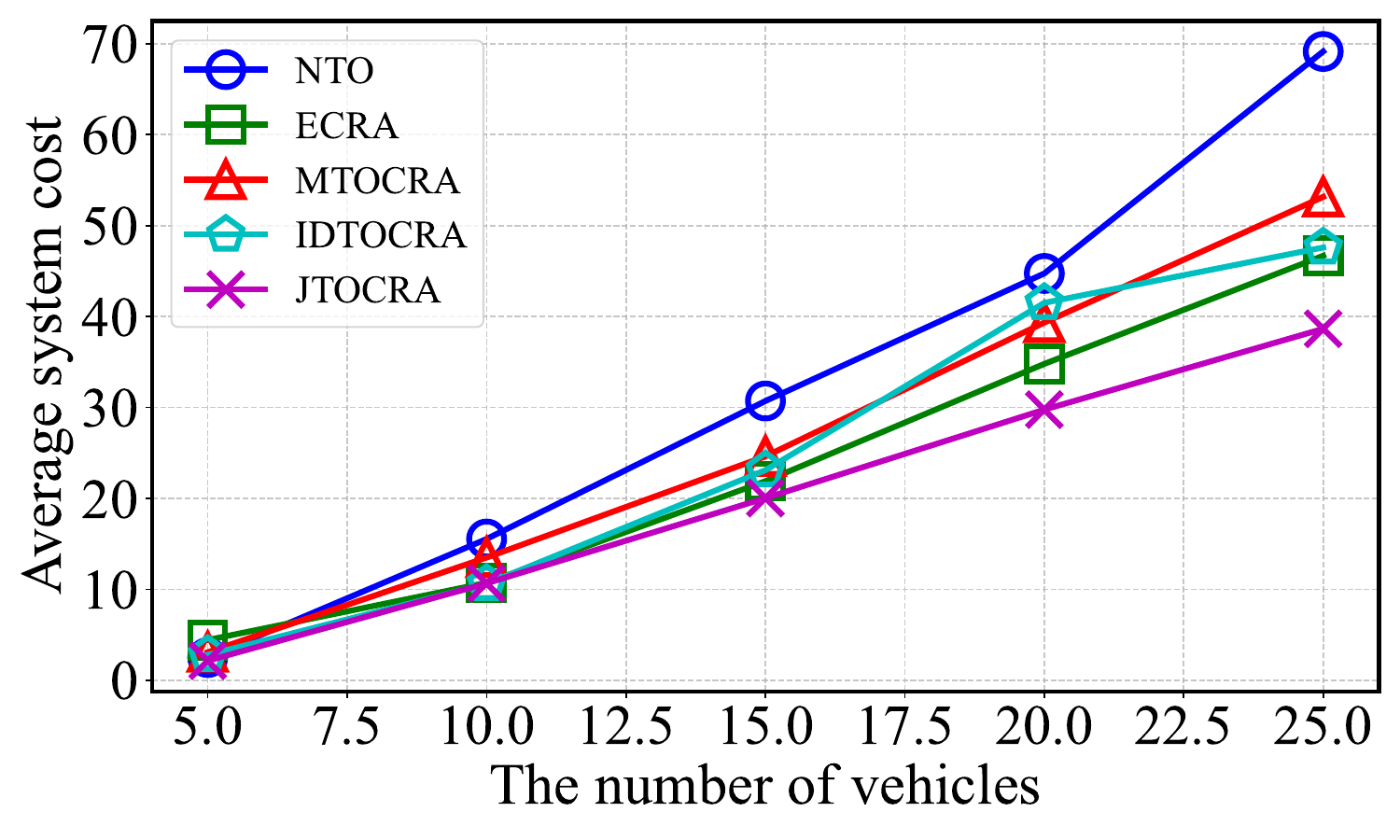}
		\end{minipage}
	}
	\subfigure[Average service delay]
	{
		\begin{minipage}[t]{0.3\linewidth}
			\centering
			\includegraphics[width=2.2in]{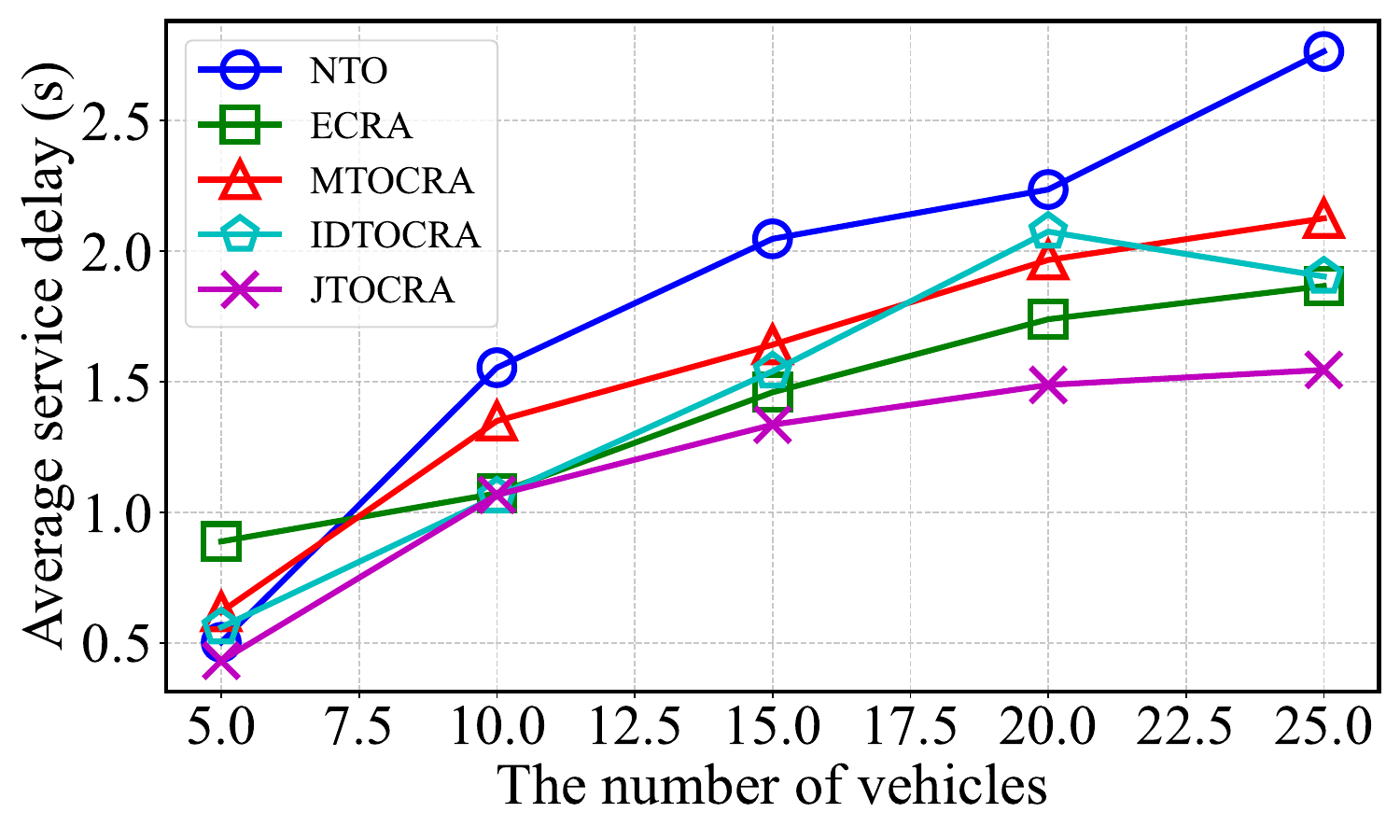}	
		\end{minipage}
	}
	\subfigure[Average energy consumption]
	{
		\begin{minipage}[t]{0.3\linewidth}
			\centering
			\includegraphics[width=2.2in]{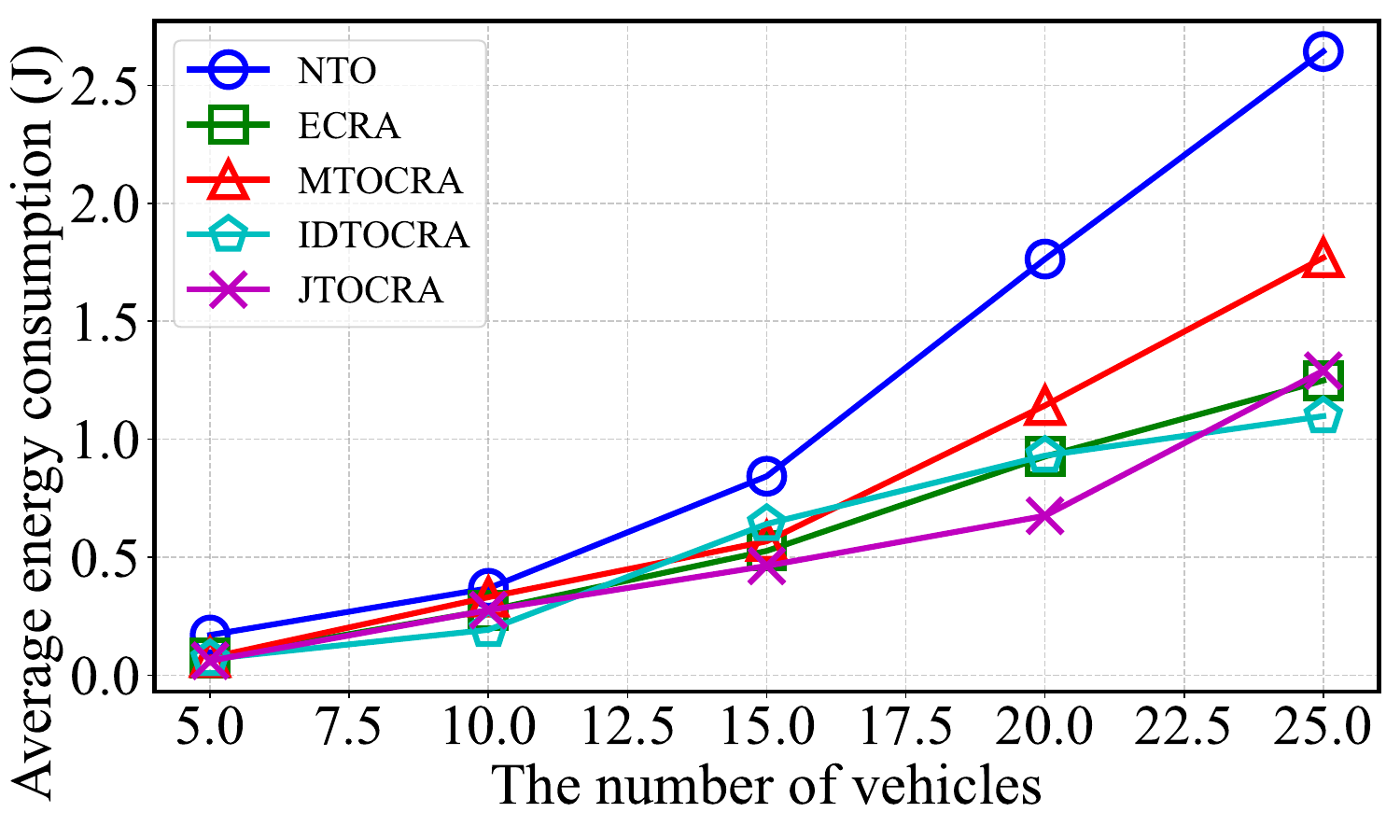}
		\end{minipage}
	}
	\centering
	\caption{System performance with the number of vehicles.}
	\label{fig_veh}
	\vspace{-1em}
\end{figure*}

\begin{figure*}[!hbt] 
	\centering
	\subfigure[Average system cost]
	{
		\begin{minipage}[t]{0.3\linewidth}
			\raggedleft
			\includegraphics[width=2.2in]{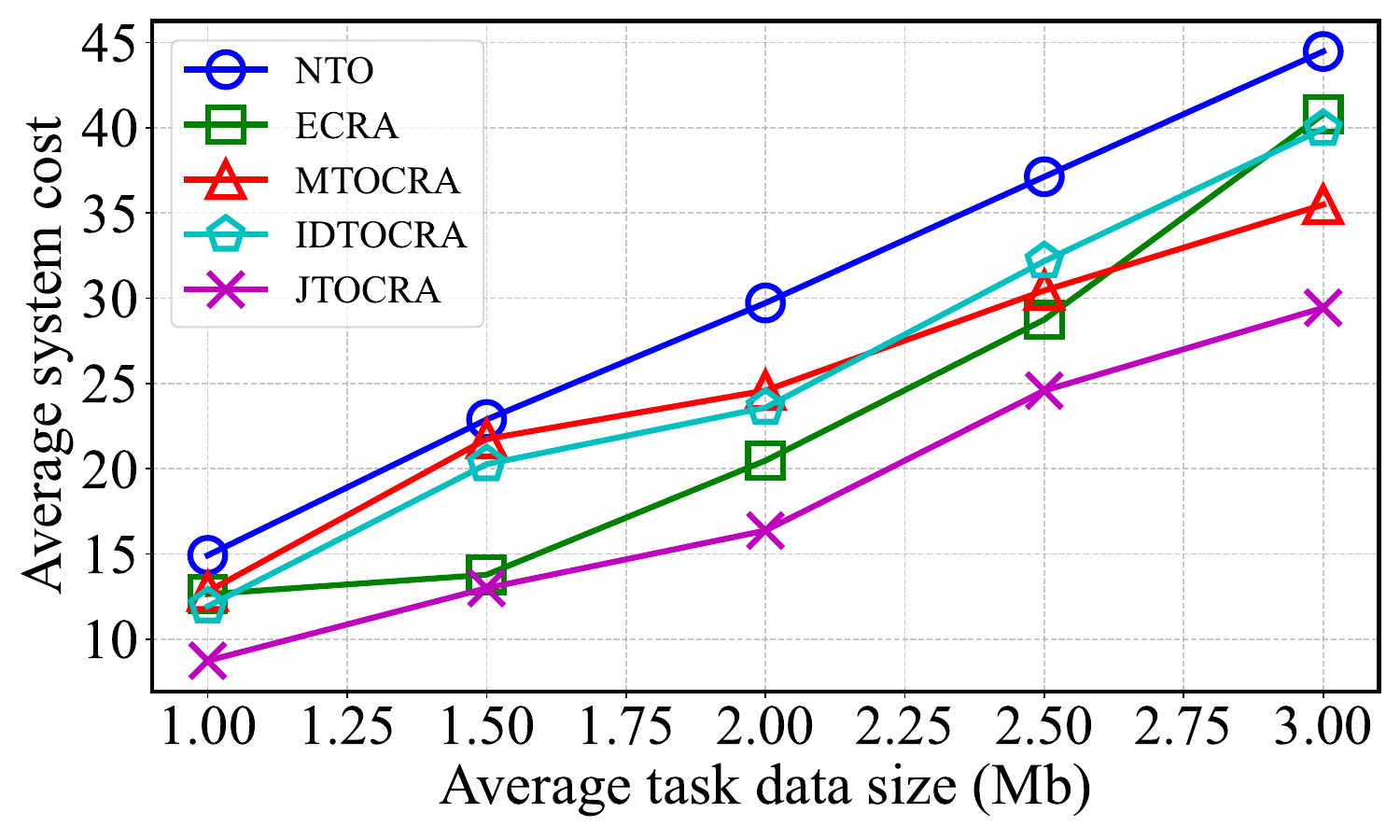}
		\end{minipage}
	}
	\subfigure[Average service delay]
	{
		\begin{minipage}[t]{0.3\linewidth}
			\centering
			\includegraphics[width=2.2in]{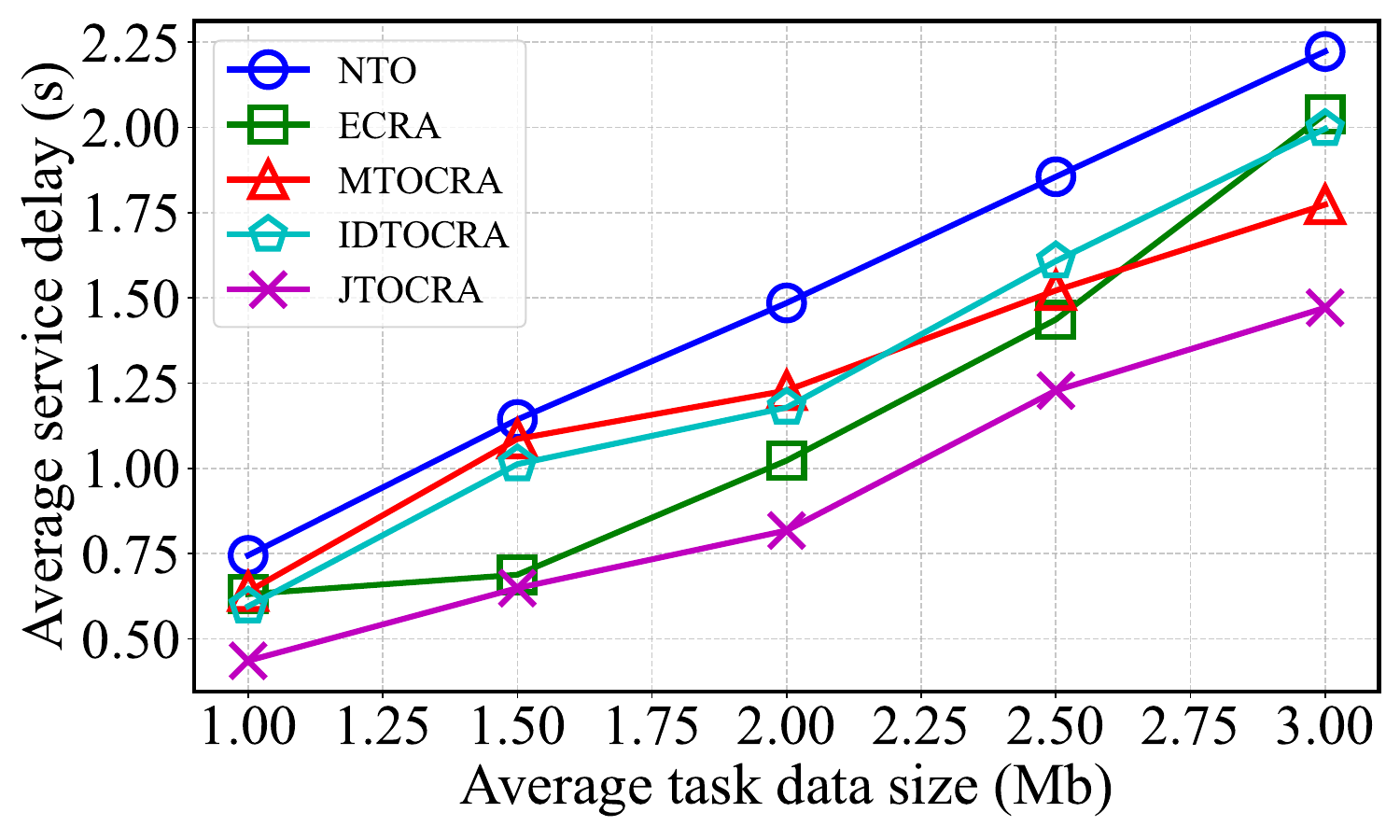}	
		\end{minipage}
	}
	\subfigure[Average energy consumption]
	{
		\begin{minipage}[t]{0.3\linewidth}
			\centering
			\includegraphics[width=2.2in]{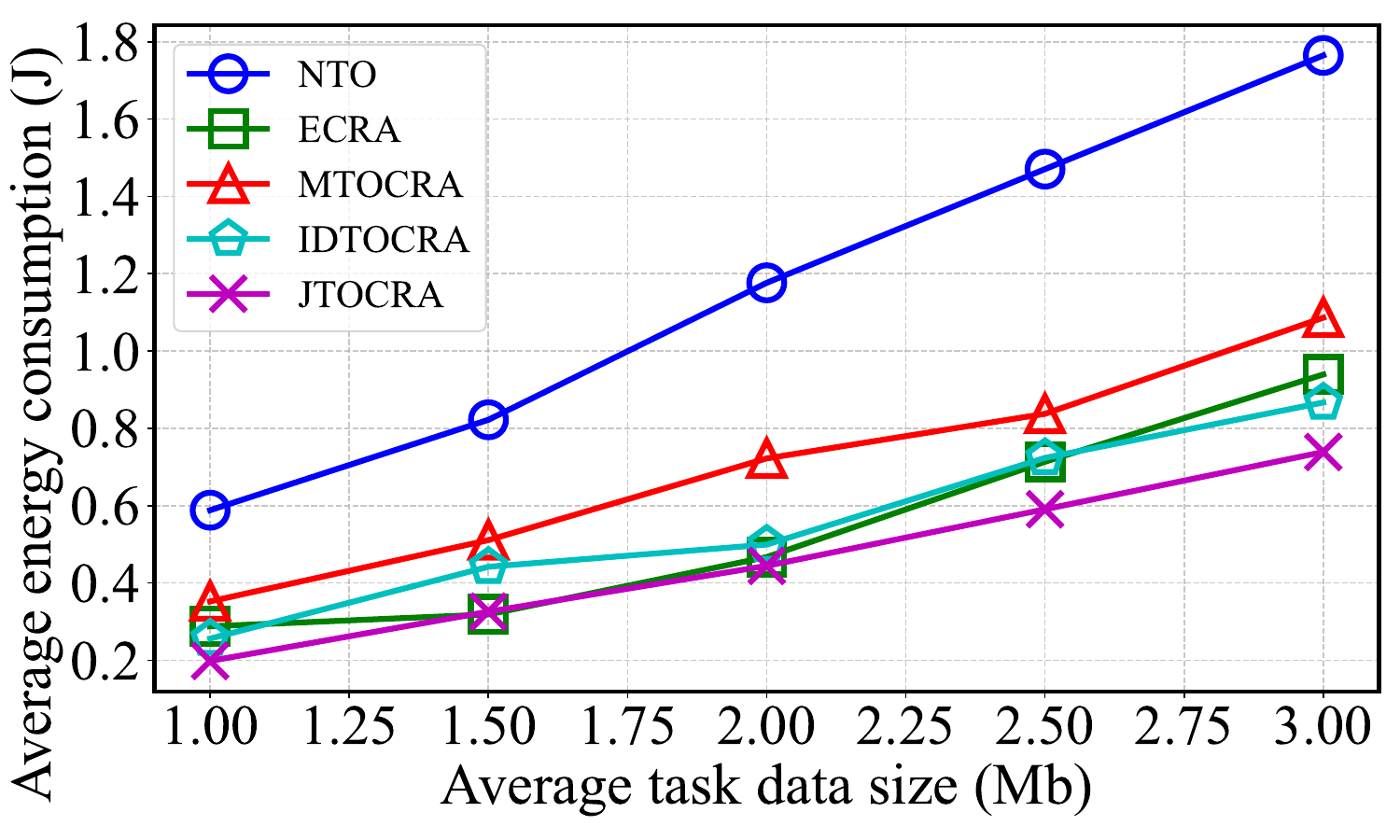}
		\end{minipage}
	}
	\centering
	\caption{System performance with task data sizes.}
	\label{fig_task}
	\vspace{-1em}
\end{figure*}

\begin{figure*}[!hbt] 
	\centering
	\subfigure[Average system cost]
	{
		\begin{minipage}[t]{0.3\linewidth}
			\raggedleft
			\includegraphics[width=2.2in]{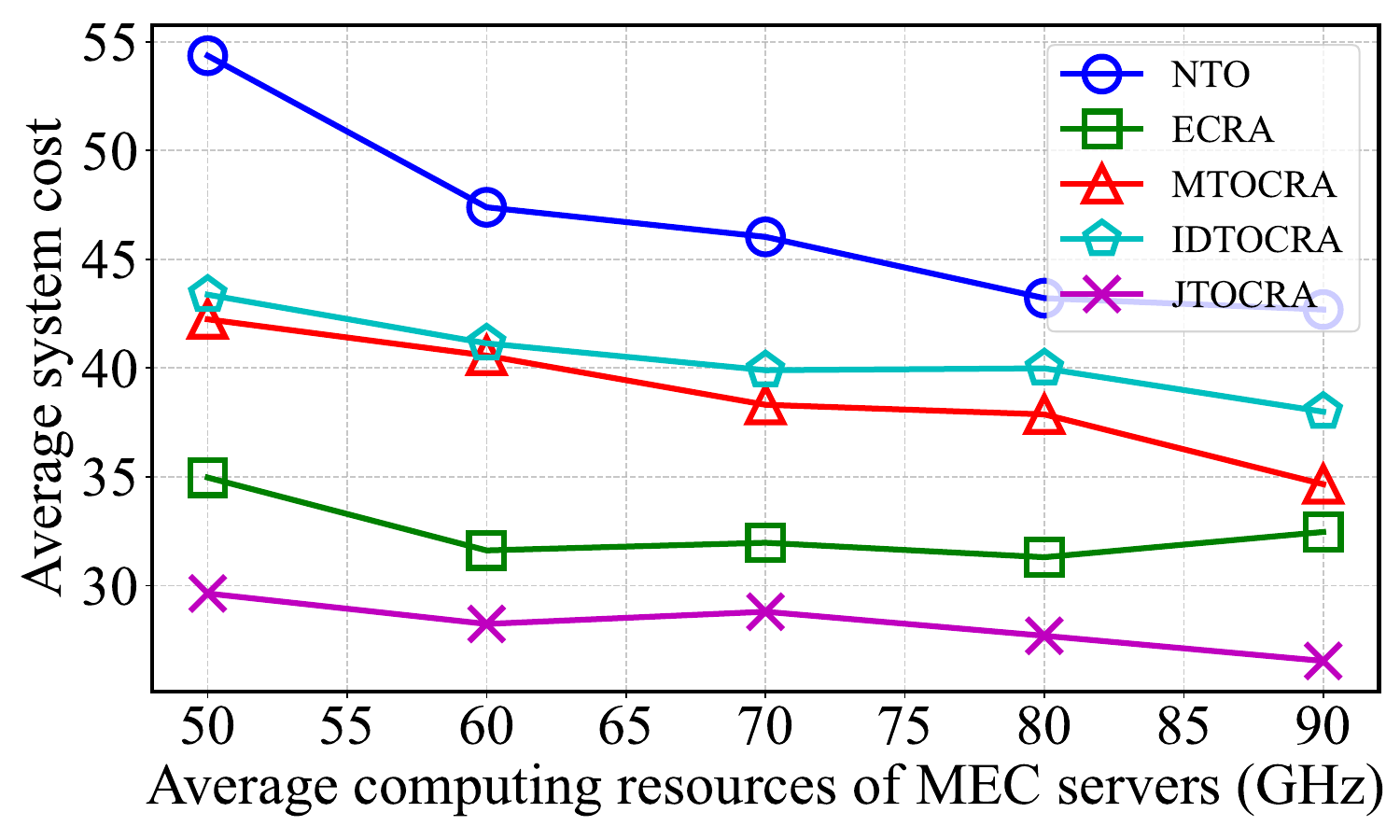}
		\end{minipage}
	}
	\subfigure[Average service delay]
	{
		\begin{minipage}[t]{0.3\linewidth}
			\centering
			\includegraphics[width=2.2in]{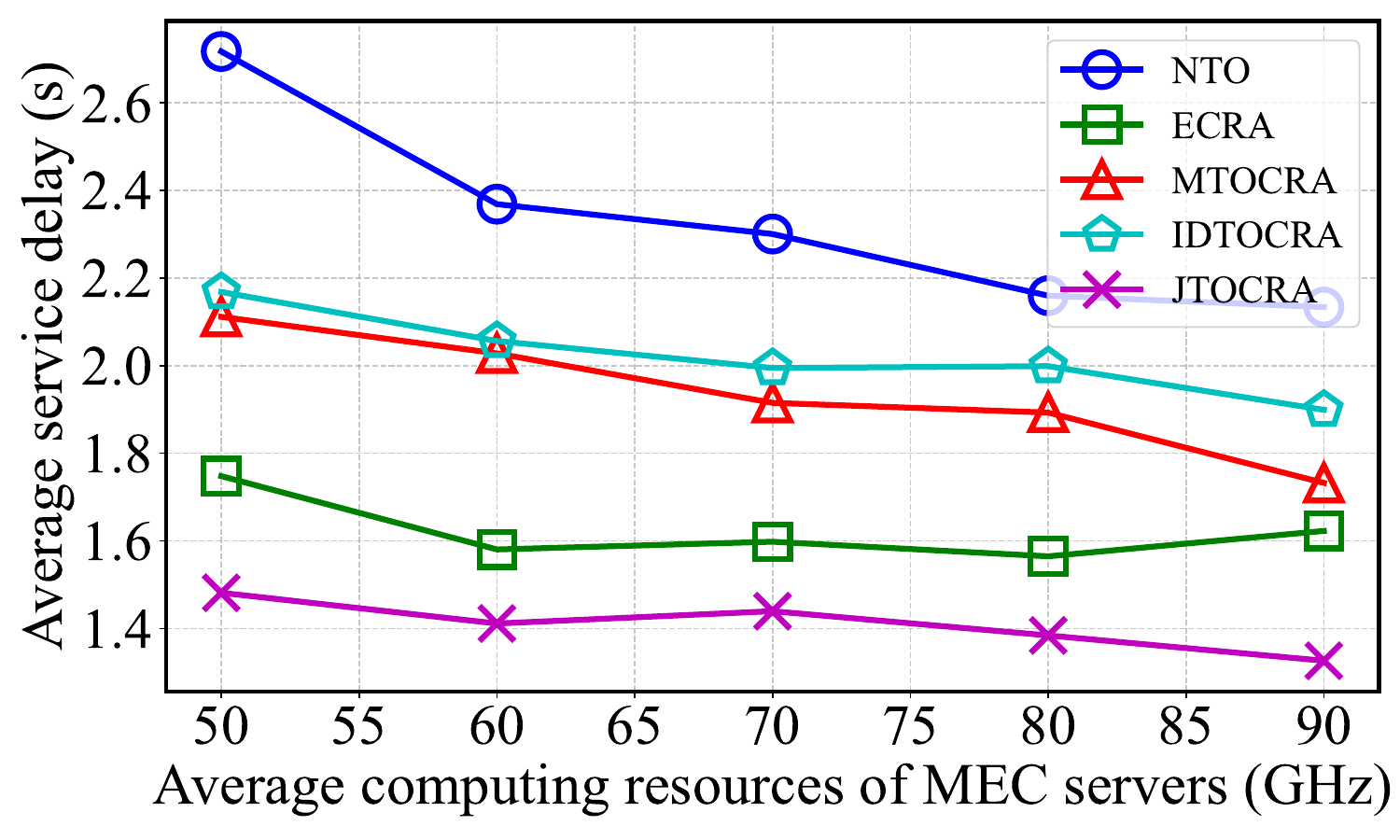}	
		\end{minipage}
	}
	\subfigure[Average energy consumption]
	{
		\begin{minipage}[t]{0.3\linewidth}
			\centering
			\includegraphics[width=2.2in]{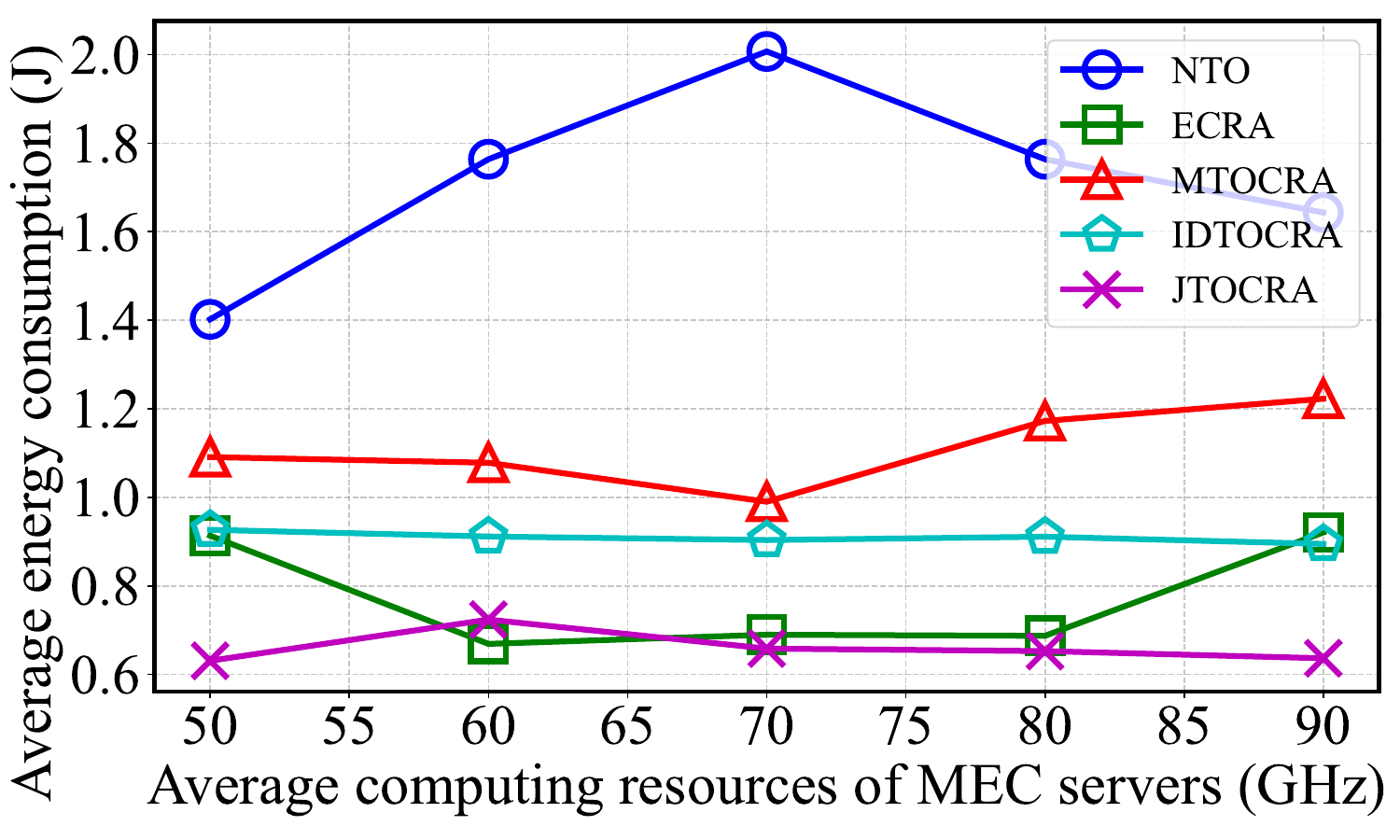}
		\end{minipage}
	}
	\centering
	\caption{System performance with average computing resources of MEC servers.}
	\label{fig_res}
	\vspace{-1.5em}
\end{figure*}

\par \textbf{Impact of Vehicle Numbers.} Figs. \ref{fig_veh}(a), \ref{fig_veh}(b), and \ref{fig_veh}(c) illustrate the impact of vehicle numbers on the performances of average system cost, average service delay, and average energy consumption, respectively. It can be observed from Fig. \ref{fig_veh} that the average system cost, average service delay, and average energy consumption of the five approaches exhibit an overall upward trend as the number of vehicles increases due to the generation of more tasks. Moreover, the proposed JTOCRA demonstrates superior performance compared to the other benchmarks, achieving performance improvements of approximately 31.55\%, 18.41\%, 24.29\%, and 16.53\% in average system cost, 31.54\%, 18.42\%, 24.29\%, and 16.53\% in average service delay, as well as 49.40\%, 10.30\%, 24.15\%, and 1.05\% in average energy consumption. The reasons are as follows. First, the inferior performance of NTO and ECRA can be attributed to the inefficiencies of the nearest offloading strategy and the average resource allocation strategy. Specifically, the nearest offloading strategy of NTO often results in workload imbalance because it fails to account for the varying requirements of tasks and the differing capabilities of MEC servers. Similarly, the average resource allocation strategy of ECRA does not consider the dynamic demands and resource availability in a fluctuating IoV environment, leading to inefficient resource utilization. As a result, both NTO and ECRA experience increased system costs, longer service delays, and higher energy consumption, especially in scenarios with a high number of vehicles. 

\par Second, the inferior performance of MTOCRA and IDTOCRA is mainly due to their limited ability to coordinate among multiple vehicles. Specifically, MTOCRA, while better at coordination, relies on policy optimization that may not capture the complex interactions in a dynamic multi-vehicle IoV environment as effectively as MADDPG. For IDTOCRA, each vehicle operates as an independent agent, leading to uncoordinated decisions that can result in resource conflicts. Notably, IDTOCRA shows a downward trend in average service delay when the number of vehicles exceeds 20. This occurs because independent decision making leads to resource competition in dense scenarios, causing more tasks to fail in offloading, which increases available resources and reduces delay. However, this decrease in delay comes at the cost of a higher task failure rate, which can negatively impact overall system performance. In conclusion, \textcolor{b}{this set of results indicates} that the proposed JTOCRA has better scalability, efficiently handling delay-sensitive and computationally intensive tasks while maintaining lower system cost, service delay, and energy consumption.

\par \textbf{Impact of Computing Resources of MEC Servers.} Figs. \ref{fig_res}(a), \ref{fig_res}(b), and \ref{fig_res}(c)
compare the impact of the average computing resources on the performances of average system cost, average service delay, and average energy consumption, respectively. It can be seen from the figure that with the increase in the computing resources of MEC servers, there is an upward trend in terms of the average system cost, average service delay, and average energy consumption. First, it can be observed from Figs. \ref{fig_res}(a) and \ref{fig_res}(b) that with the increase in the MEC computing resources, the five approaches show an overall decreasing trend in terms of the average system cost and average service delay, which can be attributed to the benefits of enhanced computing capabilities in reducing overall system overhead. 

\par Moreover, it can be seen from Fig. \ref{fig_res}(c) that the NTO shows an significant increase followed by an obvious decrease in energy consumption. In contrast, both MTOCRA and ECRA exhibit a slight initial downward trend and a subsequent upward trend, while the IDTOCRA maintains a constant tendency in the energy consumption. The main reasons for the phenomena \textcolor{b}{are} explained as follows. For NTO, with insufficient computing resources, the nearest offloading could cause task overload at some MEC servers, leading to an initial rise in energy consumption. However, as computing resources continue to increase, the overloaded MEC servers can handle tasks more efficiently. For MTOCRA and ECRA, the initial increase in MEC computing resources allows for more efficient task processing since the resource allocation strategy of ECRA and MTOCRA would not cause load imbalance compared to NTO, which results in a reduction in energy consumption. However, as computing resources continue to rise, the marginal gains from additional resources diminish, and the overhead of utilizing the computing resources starts to outweigh the benefits, leading to an increase in energy consumption. For IDTOCRA, the stability likely results from the lack of coordination of independent decision-making, leading to the less efficiency of resource utilization. 

\par Besides, compared with the other approaches, the proposed JTOCRA achieves significant superior performances in terms of average system cost, average service delay, and average energy consumption, which demonstrates that the proposed approach is able to achieve sustainable and efficient computing resource utilization.

\par \textbf{Impact of Task Data Size.} Figs. \ref{fig_task}(a), \ref{fig_task}(b), and \ref{fig_task}(c) depict the impact of the task data size on the performances of average system cost, average service delay, and average energy consumption, respectively. As shown in Fig. \ref{fig_task}, all approaches show a consistent upward tendency in terms of the average system cost, the average service delay, and the average energy consumption as the task data size increases. This trend is mainly because the larger task data size indicates heavier workload, which results in higher computational overheads, thereby leading to increased service delay and energy consumption. Moreover, with the increasing of the task data size, NTO exhibits a significant upward trend in terms of the average system cost, average service delay, and average energy consumption, consistently performing at an inferior level compared to the other approaches. This is because NTO offloads the tasks of each vehicle to the nearest MEC server without considering the varying computing capabilities and workloads of different MEC servers. This could lead to congestion and overload at certain MEC servers, further exacerbating the performances in service delay and energy consumption. In contrast, the proposed JTOCRA demonstrates superior performance, reducing the average system cost by 33.82\%, 27.83\%, 17.05\%, 26.29\%,  decreasing the average service delay by 33.81\%, 27.83\%, 17.05\%, 26.29\%, and respectively reducing the average energy consumption by 58.14\%, 21.36\%, 31.99\%, 14.77\% in the scenario with relative heavier workloads (where the average task size reaches 3 Mb). In conclusion, this set of results indicates that the proposed JTOCRA is able to adapt to the heavy-loaded scenarios with overall superior performances.

%
% Conclusion
%

\section{Conclusion}
\label{sec_conclusion}
\par In this paper, we explored the integration of task offloading and resource allocation in \textcolor{b}{an MEC}-assisted consumer IoV system. Initially, we introduced a multi-MEC-assisted consumer IoV system, and formulated an optimization problem SCMOP to minimize the system cost. Since SCMOP is an NP-hard and non-convex MNILP problem, we proposed JTOCRA to solve it. Specifically, the SCMOP was first reformulated as a Markov decision game. Then, we designed an MADDPG-based algorithm to learn the optimal decisions of task offloading and computing resource allocation. Finally, simulation results demonstrated that the proposed algorithm can not only achieve superior performance in terms of the system cost and task service delay, but also has better scalability compared with the other comparison algorithms. \textcolor{b}{The main limitation of the proposed JTOCRA is its reliance on the deployment of the BSs, which in turn depends on the terrestrial infrastructure. With the rapid deployment of low-altitude automated aerial vehicles (AAV), our future research will focus on efficient AAV-assisted offloading, air-ground cooperative resource allocation, and energy-aware AAV trajectory control in IoV systems.}.

%\section*{Acknowledgments}

%{\appendices
%\section*{Proof of the First Zonklar Equation}
%Appendix one text goes here.
% You can choose not to have a title for an appendix if you want by leaving the argument blank
%\section*{Proof of the Second Zonklar Equation}
%Appendix two text goes here.}

%section{References}

\bibliographystyle{IEEEtran}
\bibliography{references.bib}

\newpage

% \section{Biography Section}

% \begin{IEEEbiography}[{\includegraphics[width=1in,height=1.23in,clip,keepaspectratio]{figure/profile/Zemin}}]{Zemin Sun} received a BS degree in Software Engineering, an MS degree and a Ph.D degree in Computer Science and Technology from Jilin University, Changchun, China, in 2015, 2018, and 2022, respectively. Her research interests include vehicular networks, edge computing, and game theory. 
% \end{IEEEbiography}

% \begin{IEEEbiography}
% [{\includegraphics[width=1in,height=1.25in,clip,keepaspectratio]{fig1}}]{Michael Shell}
% Use $\backslash${\tt{begin\{IEEEbiography\}}} and then for the 1st argument use $\backslash${\tt{includegraphics}} to declare and link the author photo.
% Use the author name as the 3rd argument followed by the biography text.
% \end{IEEEbiography}

% \bf{If you will not include a photo:}\vspace{-33pt}
% \begin{IEEEbiographynophoto}{John Doe}
% Use $\backslash${\tt{begin\{IEEEbiographynophoto\}}} and the author name as the argument followed by the biography text.
% \end{IEEEbiographynophoto}

\vfill

\end{document}